\documentclass[prx,twocolumn,aps,epsf,showpacs,superscriptaddress,longbibliography,nofootinbib]{revtex4-2}
\usepackage[pdftex]{graphicx}

\usepackage{dcolumn}
\usepackage{bm}
\usepackage{latexsym} 
\usepackage[nointegrals]{wasysym} 
\usepackage{amsmath}
\usepackage{amssymb}
\usepackage{color}
\usepackage{array}
\newcolumntype{M}[1]{>{\centering\arraybackslash}m{#1}}
\usepackage{bbm}
\usepackage[hidelinks]{hyperref}
\hypersetup{
    colorlinks,
    linkcolor=blue,
    citecolor=blue
}
\usepackage{color}
\usepackage{array}
\usepackage{cancel}
\usepackage[dvipsnames]{xcolor}
\usepackage{float}
\usepackage[export]{adjustbox}
\usepackage{comment}
\usepackage[normalem]{ulem}

\usepackage{titlesec}
\titleformat{\subsubsection}[runin]
        {\itshape}
        {\thesubsubsection.~}
        {0.0em}
        {}
        [ -- ~]
\titlespacing*{\subsubsection}{0pt}{4pt}{0pt}
\titleformat{\paragraph}[runin]
        {\itshape}
        {}
        {0.0em}
        {}
        [ -- ~]
\titlespacing*{\paragraph}{0pt}{4pt}{0pt}

\newcommand{\<}{\langle}

\renewcommand{\>}{\rangle}

\renewcommand{\[}{\left[}
\renewcommand{\]}{\right]}

\renewcommand{\d}{\partial}

\newcommand{\Z}{\mathbb{Z}}

\newcommand{\ICS}{{\rm ICS}}
\newcommand{\F}{\mathcal{F}}
\newcommand{\TO}{\mathsf{TO}}
\newcommand{\mP}{\mathcal{P}}
\newcommand{\mH}{\mathcal{H}}
\renewcommand{\mod}{{\rm mod}~}

\definecolor{joe}{RGB}{127,0,127}

\definecolor{RW}{HTML}{5BC2E7}

\newtheorem{lemma}{Lemma}[section]

\begin{document}
\title{Floquet codes and phases in twist-defect networks}
\author{Joseph Sullivan}
\email{joseph.sullivan@ubc.ca}
\affiliation{Department of Physics and Astronomy, and Stewart Blusson Quantum Matter Institute,
University of British Columbia, Vancouver, BC, Canada V6T 1Z1}
\author{Rui Wen}
\email{wenrui1024@phas.ubc.ca}
\affiliation{Department of Physics and Astronomy, and Stewart Blusson Quantum Matter Institute,
University of British Columbia, Vancouver, BC, Canada V6T 1Z1}
\author{Andrew C. Potter}
\affiliation{Department of Physics and Astronomy, and Stewart Blusson Quantum Matter Institute,
University of British Columbia, Vancouver, BC, Canada V6T 1Z1}
\begin{abstract}
We introduce a class of models, dubbed paired twist-defect networks, that generalize the structure of Kitaev's honeycomb model for which there is a direct equivalence between: i) Floquet codes (FCs), ii) adiabatic loops of gapped Hamiltonians, and iii) unitary loops or Floquet-enriched topological orders (FETs) many-body localized phases.
This formalism allows one to apply well-characterized topological index theorems for FETs to understand the dynamics of FCs, and to rapidly assess the code properties of many FC models.
As an application, we show that the Honeycomb Floquet code of Haah and Hastings is governed by an irrational value of the chiral Floquet index, which implies a topological obstruction to forming a simple, logical boundary with the same periodicity as the bulk measurement schedule.
In addition, we construct generalizations of the Honeycomb Floquet code exhibiting arbitrary anyon-automorphism dynamics for general types of Abelian topological order. 
\end{abstract}
\maketitle

\tableofcontents

\section{Introduction}
The traditional approach to topological quantum error correction involves repeatedly measuring a set of commuting stabilizers to project the system into a topologically-ordered state that encodes quantum information non-locally. Floquet codes (FCs)~\cite{HH_dynamic_2021,Paetznick_2023,Haah_Hastings_boundaries_2022,vuillot2021planar} represent an alternative paradigm, in which a periodic schedule of non-commuting measurement rounds
results in the system continually moving through a dynamically generated sequence of instantaneous code-spaces (ICSs).

FCs can exhibit potential practical advantages. For example, the original honeycomb Floquet code (HFC) of Haah and Hastings allows one to effectively measure several-qubit stabilizers through a sequence of exclusively two-qubit measurements~\cite{HH_dynamic_2021}. The HFC builds off the structure of Kitaev's honeycomb model realizing $\Z_2$ (toric-code-type) topological order, and consists of three rounds of two-body measurements, which do not commute between rounds. After the initial three rounds are all executed, these non-commuting two-body measurements effectively determine the six-body stabilizers that measure the $\Z_2$ gauge flux through each hexagonal plaquette.
Each subsequent round projects the system into a distinct ICS's each having $\Z_2$ topological order. A notable property of this code is that the logical $e$ and $m$ loop operators (corresponding to dragging a pair of $e$ or $m$ anyon excitations around a non-contractible cycle) interchange, as $e\leftrightarrow m$ after each Floquet period. While the original HFC model was studied with periodic boundary conditions, further work has been done to flatten this out into a planar geometry~\cite{Haah_Hastings_boundaries_2022,vuillot2021planar}, as required for most physical implementations. However, the $e\leftrightarrow m$ exchanging dynamics appears to provide an obstacle to simple schemes to achieving a planar HFC~\cite{Haah_Hastings_boundaries_2022}.

While intriguing examples of FCs have been constructed in specific models \cite{Aasen2022adiabatic, brown2022}, a systematic framework for characterizing their universal structure remains elusive. 
By contrast, static topological codes can be understood by an equivalence between topological error correcting codes and topological orders of gapped ground-states of local Hamiltonians, which are governed by a well-established categorical theory of anyons~\cite{Levin_2005,Kitaev_2006,kong2014braided}. 
It is natural to ask whether a similar level of understanding of FCs could be achieved by connecting them to the well-studied topology of unitary Floquet dynamics generated by local time-dependent Hamiltonians.
Progress in this direction has recently been made by exploring connections between FCs and adiabatic loops (ALs) through the space of gapped Hamiltonians~\cite{Aasen2022adiabatic}. An AL is defined as a one-parameter family of gapped Hamiltonians $H(\theta)$ with $\theta\in S^1$, i.e. $H(\theta+2\pi)=H(\theta)$. The study of the topology of ALs has a long history starting with Thouless' famous pump~\cite{Thouless_pump}. In~\cite{Aasen2022adiabatic} it was argued that there was a close correspondence between FCs and ALs, and that, in $2d$, topology of Floquet dynamics corresponded to anyon-permuting action generalizing the $e\leftrightarrow m$ exchanging dynamics of the HFC.

Here, we seek to develop further connections between FCs and and non-equilibrium dynamical topological phases of unitary dynamics, that are governed by well-understood topological invariants~\cite{Gross_2012,rudner2013anomalous,Else_2016,Po_2016,Potter_2017,Roy_2017,Po_2017,Fidkowski_2019,Duschatko_2018,zhang2021classification,zhang2022bulk}.
Specifically, we explore the relationship between topological FCs arising from non-unitary dynamics driven by a time-periodic schedule of local measurements, and unitary Floquet enriched topological orders (FETs) arising from unitary evolution, $U(t) = \mathcal{T}e^{-i\int_0^t H(s)ds}$ generated by a local, time-periodic Hamiltonian $H(t+1)=H(t)$. 

The topology of unitary dynamics is defined for unitary loops satisfying $U(t=n)=\mathbbm{1}$ for some non-zero integer $n$. 
Under appropriate conditions, unitary loops (ULs) can further be extended into stable many-body localized phases with eigenstate topological order~\cite{Harper_2020} that is modulated by topologically non-trivial micro-motion within each Floquet period.
The topology of unitary FETs with Abelian topological order in $2d$ is well-studied~\cite{Potter_2017, Po_2017}.
In $2d$, unitary loops and Floquet-MBL phases are governed by: i) emergent dynamical symmetries that permute the topological charge of anyon excitations during each period, and ii) a chiral Floquet (CF) index $\nu(U)$. 
Aspects i) and ii) are not entirely independent, rather, as the anyon permuting dynamics may constrain the allowed values of the CF index.
Floquet phases without topological order have rational values of $\nu(U)\in \mathbb{Q}$, whereas for FETs with Abelian topological order the index can take irrational ``radical"-valued indices: $\nu(U) \in \sqrt{\mathbb{Q}}$~\cite{Po_2017}. A radical-valued CF index is necessarily tied to a non-trivial permutation of bulk anyons during each Floquet period -- an anyon ``time-crystal".
Physically, $\log \nu(U)$ characterizes the amount of quantum information transported chirally along the boundary during each period~\cite{Duschatko_2018}. Additionally, there is a bulk-boundary correspondence  linking a radical value of the CF index, to a non-trivial anyon permuting dynamics of the bulk during each period.
\footnote{More precisely a radical chiral index implies a non-trivial bulk automorphism but the reverse need not be true. This is because the chiral index is set by the quantum dimension of the defect associated with the automorphism, which can be rational. In $\mathbb{Z}_4$ toric code, for example, the chiral index associated with the $e-m$ exchanging automorphism is $\nu = \sqrt{4}=2$. }

In this paper, we define a class of models with Abelian topological order, which we dub paired defect networks, for which there is a (stable) equivalence between discrete-time measurement dynamics of FCs, and continuously parameterized loops of gapped Hamiltonians, unitaries, and MBL FETs that pass through the same sequence of ICSs as the FC.
Here, by stable equivalence, we mean that the continuous loops can be defined from a FC up to stacking with some invertible topological phase (we define the notion of invertibility for dynamical phases below).
This equivalence enables us to port well-established topological index theorems and bulk-boundary correspondences for unitary FET dynamics, to establish constraints on the less-well-understood area of FCs.

As an illustrative example, we show how the honeycomb FC of Haah and Hastings~\cite{HH_dynamic_2021}, can be lifted to a unitary circuit, $U$ that has a radical CF index: $\nu(U) = \sqrt{2}\mathbb{Q}$.
We show that this implies dynamical anomaly constraints on the possible edge terminations of the HFC Floquet code with open boundaries. Specifically, we show that a gapped boundary of the Honeycomb FC is only possible if one explicitly doubles the Floquet periodicity for the boundary measurement schedule.  

We then generalize the structure of the spin-1/2 Honeycomb codes to a general family of models that we refer to as paired twist defect networks. We construct Floquet codes and phases from honeycomb networks of twist defects whose dynamics implements a general anyon automorphism, generalizing the $e\leftrightarrow m$ exchanging dynamics of the HFC with $\Z_2$ topological order to any Abelian, non-chiral topological orders. Here, by automorphism, we mean a permutation of anyons that preserves the topological properties (self- and mutual- braiding statistics, fusion properties, etc...). We describe both a general construction of these generalized HFCs and an explicit family of lattice models with only nearest-neighbor measurements for a restricted class of twist defects (those corresponding to an order-two anyon automorphism).

In addition to constructing a large class of new FC examples, we expect that our general formalism will be useful in designing new FCs and algorithmically assessing their code properties.

\section{Turning the honeycomb Floquet code into a loop}
To motivate the definition of the general paired twist-defect networks, we begin by reviewing a concrete example: the honeycomb Floquet Code (HFC) introduced by Haah and Hastings~\cite{HH_dynamic_2021}. After briefly reviewing the HFC construction, we then construct a unitary circuit that produces dynamics equivalent (in a sense defined below) to the measurement-only dynamics of the HFC. We show that this unitary circuit has an irrational CF index, $\nu(U) = \sqrt{2}\mathbb{Q}$. 
From here, we define a class of systems that generalize the structure of the HFC, discuss equivalence between FCs and Hamiltonian or unitary loops, and discuss implications of the irrational CF index for building planar versions of the HFC.

\begin{figure}
\includegraphics[width= \columnwidth]{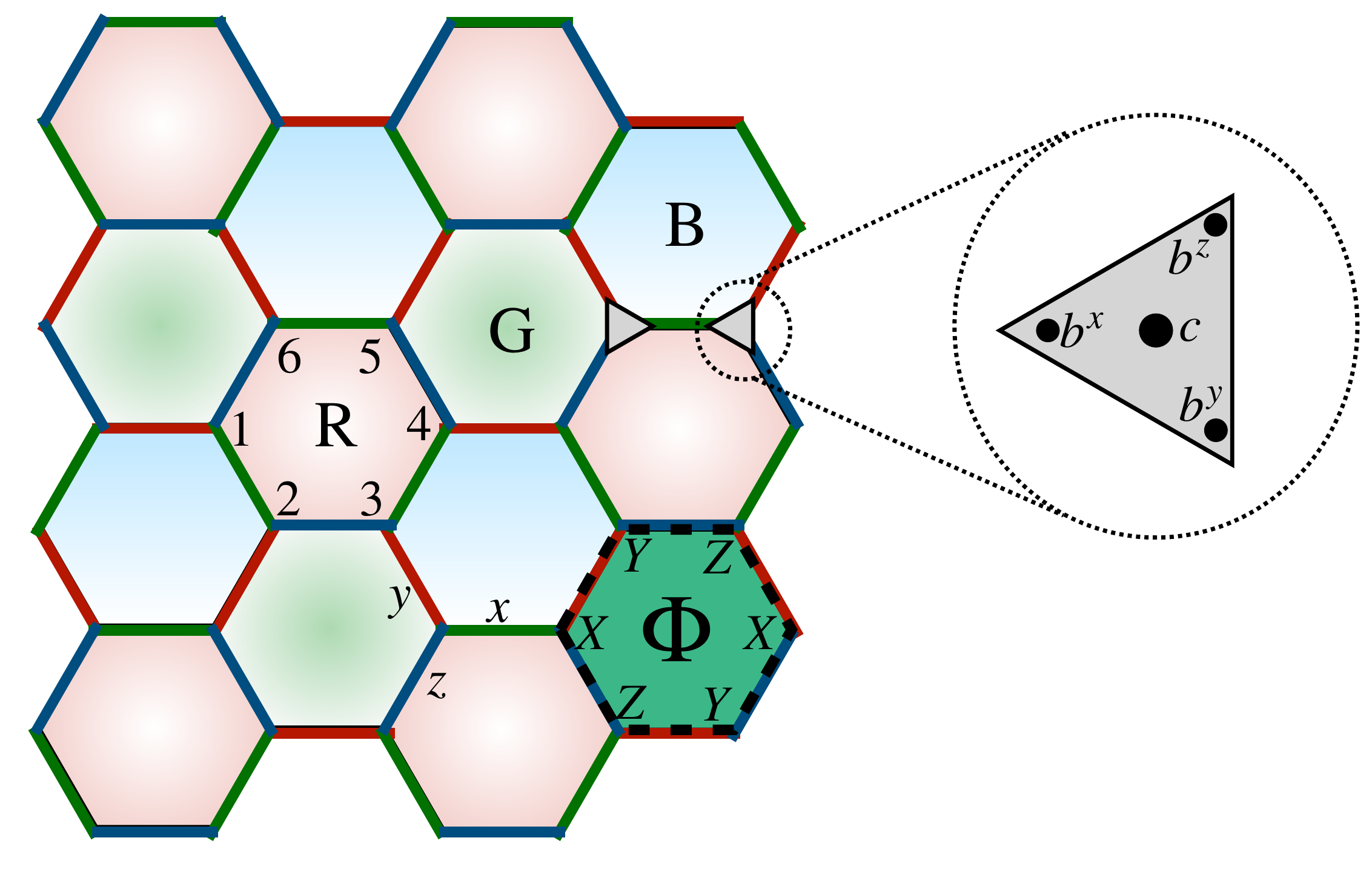}
    \caption{{\bf Honeycomb Floquet Code (HFC) -- } the HFC is defined on a three-colored honeycomb, with a spin-$1/2$ on each site/vertex. The plaquettes are labeled red (R, light-centered), green (G, dark centered), and blue (B, vertical gradient), the three distinct bonds/edges are labeled $x$, $y$, and $z$, and the sites of each plaquette are labeled from $1$ to $6$ as shown. The gray triangle depicts the standard mapping of spins to Majorana fermions, $c, b^{x,y,z}$, these can be grouped into central fermions, $c$, on each site, and $\Z_2$ gauge links $u_{ij}$ on each edge as explained in the text.}
    \label{fig:HFC}
\end{figure}

\subsection{Review of the HFC}
\label{sec:Review of the HFC}
The HFC consists of a periodically-repeating sequence of three measurement rounds acting on qubits arranged on the sites of a honeycomb.
To define the model, label the three distinct orientations of bonds on the honeycombs by label $\alpha\in \{x,y,z\}$ as shown in Fig.~\ref{fig:HFC}. Introduce a three-coloring of plaquettes red (R), green (G), and blue (B). In addition to the bond-orientation ($x,y,z$) labels, also label bonds that connect $R,G,B$ plaquettes with $R,G,B$ respectively. For future convenience, we also number the sites around each hexagonal plaquette, $p$ by $1\dots 6$ starting from the middle-left corner and going around in the counter-clockwise direction. 

Following the notation of Kitaev's honeycomb model~\cite{Kitaev_2006}, define the bond-operators $P_{ij} = \sigma^\alpha_i\sigma^\alpha_j$ where $\sigma^{\alpha=x,y,z}$ are spin-1/2 Pauli operators for direction $\alpha$ that coincides with the $x,y,z$ label of bond $ij$. Further, for each site $i$, introduce the Majorana fermion operators $c_i,b^\alpha_i$ related to the spin operators by $\sigma^\alpha = ib^\alpha_i c_i$. The physical spin model is recovered by projecting $ic_ib^x_ib^y_ib^z_i=1$ on each site. It is convenient to repackage these operators as Majorana ``defects" $c_i$ on each site, and $\Z_2$ gauge connections $u_{ij} = ib^\alpha_{\<ij\>_1} b^\alpha_{\<ij\>_2}$ on each bond. Here, we choose an orientation for each bond, $\<ij\>$, and $\<ij\>_{1,2}$ label the start, end sites of the bond respectively (in this way $u_{ij}=u_{ji}$ and one may henceforth ignore the bond orientations). In this representation $P_{ij} = ic_iu_{ij}c_j$ are simply interpreted as the (gauge-invariant version of the) fermion parity of the pair of Majorana defects $c_i,c_j$.

The HFC~\cite{HH_dynamic_2021} then consists of a repeating sequence of three rounds of measurements: in round $1,2,3$, $P_{ij}$ are measured, or ``checked", for each $R,G,B$ bonds respectively. We refer to these measured bond operators as parity checks or simply ``checks"\footnote{In the subsystem code literature $P_{ij}$ are often referred to as gauge operators or gauge checks; a notation that is ripe for confusion with standard gauge theory formulation of topological orders.}. While the measurement schedule repeats periodically, the checks in round $r$ do not commute with those in round $r+1$, and hence the measurement outcomes are random and generically non-repeating. However, after three rounds, the products of checks $P_{ij}$ around any hexagon $P$ (and hence also any contractible loop),  given by $\Phi_P = \prod_{\<ij\> \in \hexagon_p} P_{ij}$, are determined. $\Phi_P$ then commute with all future measurement checks, to form persistent stabilizers of the HFC. In the gauged-fermion language, the persistent stabilizers are simply the $\Z_2$ gauge flux $\Phi_p = \prod_{\hexagon_p} u_{ij} = \pm 1$. Crucial to the performance of the code, the measurement sequence should be chosen to avoid measuring the flux through non-contractible loops, which represent logical operators.

After each measurement round, the most-recently-measured parity checks and persistent gauge-flux stabilizers together form the stabilizers of a $\Z_2$ topological order that is topologically equivalent to a toric code. This is referred to as an instantaneous code space (ICS). The three different ICSs after the $R,G,B$ measurement rounds are respectively labeled $\ICS_{R,G,B}$. Each ICS is represented by a distinct, but topologically equivalent, pairing of the Majorana defects, immersed in a fixed gauge-flux configuration dictated by the persistent stabilizer values.
The states of each ICS can be labeled by configurations of point-like anyon excitations with topological charges (a.k.a. topological super-selection sectors), $f,m, \text{ and }e=m\times f$. We associate a flipped bond stabilizer $P_{ij}=-1$ with a fermion ($f$) excitation.  Since the bond-stabilizers are on different ($R,G,B$) bonds during each round, the resulting $e,m$ excitation labeling also depends on the round. For $\ICS_R$, we label the $\Z_2$ gauge fluxes, $\Phi_p=-1$ on the $p\in R$ plaquettes as $m$ excitations, and those on the $B,G$ plaquettes as $e$ excitations. 
To understand this color-dependent labeling, note that the local operator $S^y_4=ic_4b^y_4$ in Fig.~\ref{fig:HFC} changes $u_{45}\rightarrow -u_{45}$, i.e. creates fluxes on plaquettes labeled $R$ and $G$ and also creates a fermion on the red bond touching site $4$. This shows that it would not be consistent to label all gauge fluxes as $m$ regardless of color. In our labeling convention this simply corresponds to the fusion rule $1=e\times m\times f$. 
Similarly for $\ICS_G$ ($\ICS_B$) denote the $G$ ($B$) fluxes as $m$ excittions and the $R,B$ ($R,G$) plaquette fluxes as $e$ excitations. 

The logical operators of each ICS can also be labeled by anyon types $e,m,f=e\times m$. 
At a coarse grained level, each type of logical operator, $a$, can be viewed as creating an $a$ particle/anti-particle pair, dragging them around a non-contractible cycle, and then annihilating them.
These loop operators can be decomposed into a product of many short line segments whose ends create $a/\bar{a}$ anyon/anti-anyon pairs.
Representative logical operators for the HFC on a torus are shown in Fig.~\ref{fig:majorana logical}.
The $f$-loops through a cycle indicate the $\Z_2$ gauge flux $\prod_{ij}u_{ij}$ through any co-threaded cycle, and are persistent logical operators shared by each ICS. The $e,m$ loops of round $r$ are not persistent operators, but may be augmented by multiplying by parity checks $P_{ij}$ of round $r$ to become logical operators of the next round, $r+1$. 
In this way, the measurement dynamics map each logical operator from ${\rm ICS}_r$ to those of ${\rm ICS}_{r+1\mod 2}$. A key property of the HFC is that, after three cycles, the $e$ and $m$ loops are interchanged.

\subsection{Lifting the measurement-only dynamics to a unitary loop\label{sec: majorana lift}}
Many of the dynamical features of the HFC, especially the $e\leftrightarrow m$ exchanging property, are also found in the unitary dynamics of Honeycomb models of Floquet enriched topological orders (FETs)~\cite{Fidkowski_2019,Po_2017,Potter_2017}. It is natural to ask whether these features are related. 
To establish a connection between the measurement-only FC dynamics and the unitary FET dynamics, we design a sequence of unitary circuit evolutions, $U_{1,2,3} = e^{-iH_{1,2,3}}$, that maps between the ICSs of the $R,G,B$ measurement rounds of the HFC. In other words, $U_{1,2,3}$ will respectively map each state of $\rm ICS_{R,G,B}$ to one in $\rm ICS_{G,B,R}$. A closely-realted circuit structure, dubbed Kramers-Wannier circuits, was introduced in~\cite{Aasen2022adiabatic}, and used to construct an adiabatic path through the space of gapped Hamiltonians. Here, we focus instead on the non-equilibrium Floquet dynamics of the entire spectrum of excited states generated by repeated application of this circuit, which allows us to connect to well-established topological indices for unitary loops and quantum cellular automata~\cite{Gross_2012,rudner2013anomalous,Else_2016,Po_2016,Potter_2017,Roy_2017,Po_2017,Fidkowski_2019,Duschatko_2018,zhang2021classification,zhang2022bulk}.

\begin{figure}
    \centering
\includegraphics[width= \columnwidth]{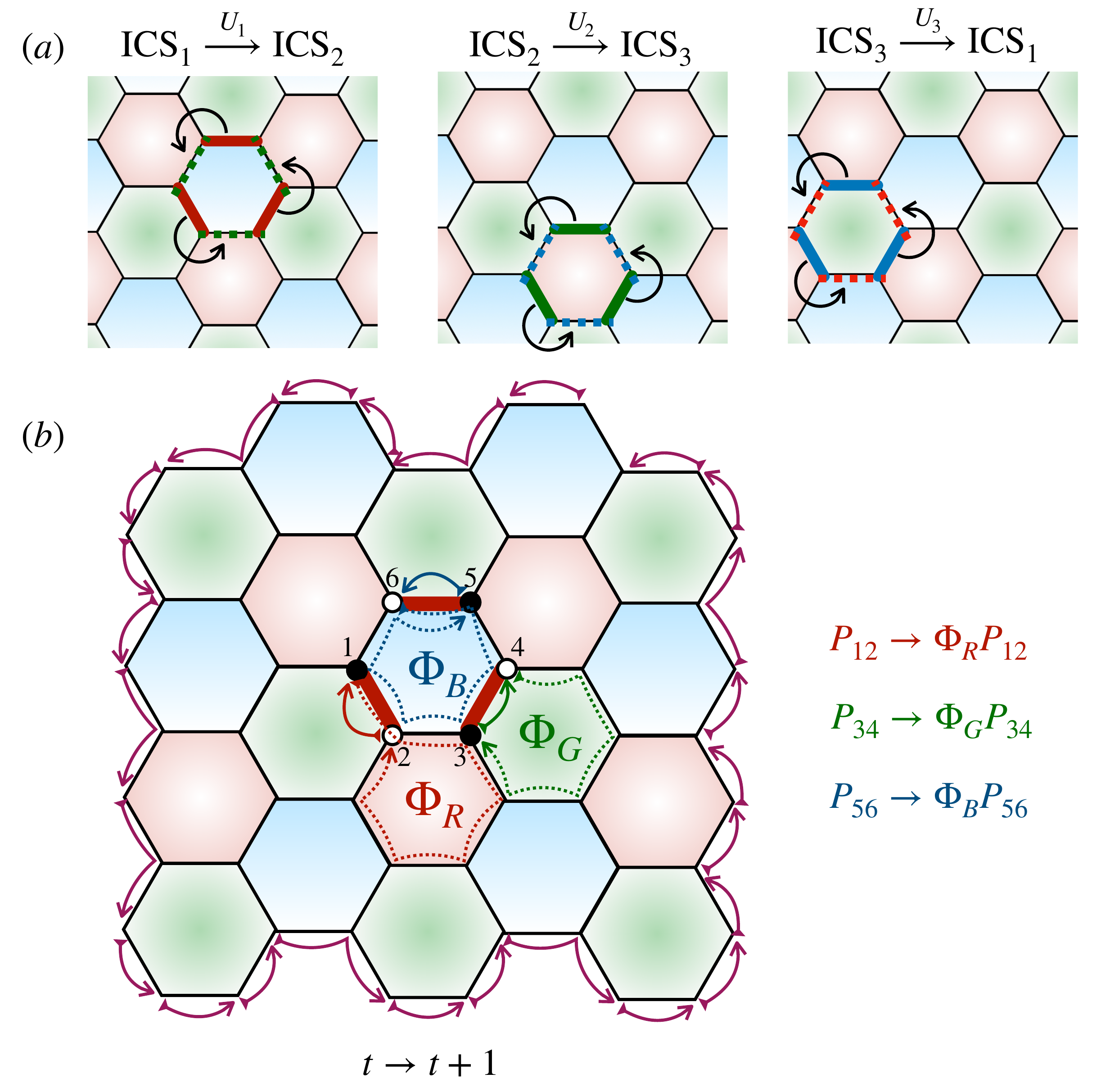}
    \caption{{\bf Dynamics of the unitarized Honeycomb Floquet Code -- } (a) Floquet unitary dynamics made up of a sequence of unitaries, $U_{1,2,3}$ that map between the ICS's of the HFC. Step $1,2,3$ consists of a counter-clockwise rotation around the $B,R,G$-color plaquettes respectively. (b) The resulting evolution of the Majorana operators $c_i$ and bond-parities $P_{ij}$ after one period: the bulk Majoranas swap such that each $P_{ij}$ evolves to itself up to a phase that depends on the gauge flux through an adjacent plaquette. The edge Majoranas evolve in a large open orbit, signalling a radical chiral Floquet index, $\nu(U_3U_2U_1)=\sqrt{2}$.}
   
    \label{fig:Unitary action}
\end{figure}

To construct the sequence of $U_a$'s, note that the parity-checks for the $R$ and $G$ measurement rounds of the HFC differ by a $60^\circ$ rotation of each $B$ plaquette, and the transitions $G\rightarrow B$ and $B\rightarrow R$ can be similarly accomplished by rotating the $R$ and $G$ plaquettes respectively. To implement this action with a unitary, we introduce local unitary operators that cyclically permute the Majorana operators counterclockwise about a given hexagonal plaquette $P$:
\begin{align} 
C_p &= B_{12}B_{23}B_{34}B_{45}B_{56},
\end{align} 
where $B_{j,i}= 
e^{\frac{\pi}{4}c_ju_{ij}c_i}$ braids Majorana mode $c_i$ around $c_j$: 
\begin{align}
B_{j,i}^\dagger \begin{pmatrix} c_i \\ c_j \end{pmatrix}B_{j,i} &= 
u_{ij}\begin{pmatrix} -c_j \\ +c_i \end{pmatrix}
\end{align}
Referring to the numbering convention of sites on each plaquette shown in Fig.~\ref{fig:Unitary action}, this cyclic permutation operator, $C_p$, has the following action on the bond-parity checks: 
\begin{align}
C_p: 
\begin{cases} P_{i,i+1}\rightarrow P_{i+1,i+2} & i\neq 5 \\ P_{5,6}\rightarrow -\Phi_p P_{12} &  \\  \end{cases}
\end{align}
where we take site numbers $i$ modulo $6$. In other words, each of the bond parity checks is cyclically permuted around the edge of the hexagon. In addition, one bond $P_{6,5}$ parity check picks up a phase equal to the gauge flux, $\Phi_p$ through the $p$; this will be crucial to recovering the $e\rightarrow m$ exchanging dynamics of the HFC.

We then define: 
\begin{align} 
U_{1,2,3} = \prod_{p\in B,R,G} C_p.
\label{eq:U123}
\end{align}
which can each be generated by local Hamiltonians $H_{1,2,3} = -i\log U_{1,2,3}$. Then, we define the continuous-time unitary Floquet evolution:
\begin{align}
    U(t) &= \mathcal{T}e^{-i\int_0^t H(s)ds} \nonumber\\
    H(t) &= 3\begin{cases}
        H_1 & 0\leq t< 1/3 \\
        H_2 & 1/3\leq t< 2/3 \\
        H_3 & 2/3\leq t< 1
    \end{cases}
\end{align}
where $\mathcal{T}$ denotes time-ordering, and we have set the Floquet period to $1$ for convenience. 

\begin{figure*}
    \centering
    \includegraphics[width = 1.7\columnwidth]{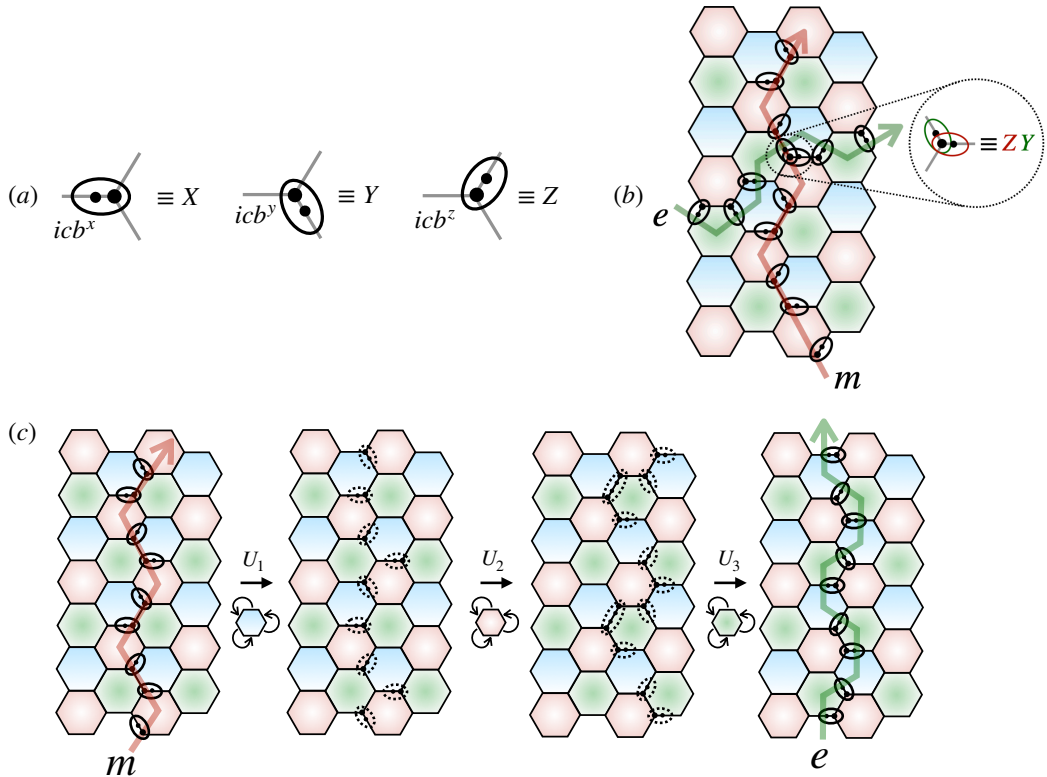}
    \caption{{\bf Logical operators of the Honeycomb Floquet Code (HFC) on a torus} (a) The onsite Pauli operator $\vec{\sigma}$ can be represented in the Majorana representation by short bubbles which enclose $c$ and $b^\sigma$, where each bubble indicates the fermion parity of the pair of enclosed Majorana defects. This provides a pictorially simple way of expressing and evolving the logical operators.
    (b) The $m$ (upward red line/arrow) and $e$ (rightward green line/arrow) logical operators can be built from the bubble operators. $e$ and $m$ logical operators around distinct cycles of the torus cross at an odd number of points, where they anticommute. (c) The evolution of an $m$ string under the Floquet unitary in Eq \ref{eq:U123}. After one cycle an $m$ logical evolves into an $e$ logical.}
    \label{fig:majorana logical}
\end{figure*}

\paragraph{Bulk dynamics} Let us first examine the bulk dynamics of this unitary model on a closed graph without a boundary. Consider starting from a state in $\ICS_R$ with a given set of anyon excitations, and applying $U(t=1) = U_3U_2U_1$. The gauge fluxes, $\Phi_p$ commute with unitaries in each step, and remain invariant under the evolution. The bond parities switch according to:
\begin{align}
    P_{12}&\rightarrow \Phi_R P_{12} \nonumber\\
    P_{34}&\rightarrow \Phi_G P_{34} \nonumber\\
    P_{56}&\rightarrow \Phi_B P_{56}, 
\end{align}
where the site and plaquette labeling is indicated in Fig.~\ref{fig:Unitary action}.
Crucially, if there is a gauge flux on plaquette $p$ after one period, one adjacent bond parity of flips, i.e. during each period each flux binds a fermion, permuting $e$ and $m$ excitations. 
The unitary dynamics also exchanges the $e$ and $m$ logical operators of the code as shown in Fig~\ref{fig:majorana logical}b.
Evolving for two periods preserves each stabilizer of $\ICS_R$ as well as the logical operators, and hence the model indeed satisfies the desired unitary loop property. 

\paragraph{Edge dynamics and topological invariant}
We can introduce an edge into the unitary model by considering the system on an open sub region, $A$ of the honeycomb and applying $C_p$ only to plaquettes $p$ that lie completely within $A$. To visualize the resulting dynamics, note that, acting on the Majorana operators $c_i$ the idealized unitaries defined above simply ``hop" the Majoranas between sites, attaching a gauge string $\prod_{ij} u_{ij}$ as they move. The resulting pattern of motion is shown in Fig.~\ref{fig:Unitary action}. Majorana modes in the bulk are locally swapped along $R$ bonds, whereas those on the boundary undergo long chiral loops encircling the system. 

One can verify by inspection that one Majorana mode is translated across each boundary bond per Floquet period. This is the hallmark of a radical chiral Floquet (CF) phase~\cite{Po_2017,Fidkowski_2019}, which are characterized by an irrational unitary CF index $\nu[U]= \sqrt{2}\mathbb{Q}$~\cite{Po_2017, Fidkowski_2019}. Since the gauge field dynamics of $u_{ij}$ is trivial in this model, we can fix the values of $u_{ij}$ and consider only the residual Majorana degrees of freedom $c_i$ in this gauge fixed background. It is then straightforward to confirm that the Majorana translation dynamics of the edge of this model result in a radical value of the CF index: $\nu[U] = \sqrt{2}$. For details we refer the reader to Appendix~\ref{chiral index} where we evaluate this index by directly evaluating it via the formula defined in~\cite{Fidkowski_2019}.

\paragraph{(Non)uniqueness of the unitarized model}
We note that the two conditions that i) the sequence of unitaries produce the same sequence of ICS's as the HFC, and ii) the unitary evolution forms a unitary loop, do not uniquely specify the model. However, previous rigorous results~\cite{Fidkowski_2019} show that the CF index exhaustively classifies unitary loops of interacting fermion systems, and establish a bulk-boundary correspondence in which the $e\rightarrow m$ exchanging bulk dynamics always accompanies an irrational CF index $\nu\sim \sqrt{2}\mathbb{Q}$. Under stacking of unitaries the CF index multiplies: $\nu(U_1\otimes U_2)=\nu(U_1U_2) = \nu(U_1)\nu(U_2)$. Hence, any other way of lifting the HFC to a unitary loop would at most differ from this model by stacking with an ``invertible" rational CF phase that does not affect the quantum information storage properties of the model. For example, reversing the orientation of the action of $C_p\rightarrow C_p^\dagger$ in each step produces a model with analogous properties, but with the inverse value of the unitary CF index $\nu = 1/\sqrt{2} = \frac12 \sqrt{2}$ which differs from $\nu(U)$ for the counterclockwise model by stacking with a rational CF phase with $\nu = \frac 12$. 
We will show below, for a large class of FCs that generalize the spin-1/2 HFC structure, that the unitary loop version of the FC is defined up to stacking with an invertible (i.e. with rational CF index) Floquet topological phase.

\paragraph{Extension to MBL-protected Floquet enriched topological (MBL-FET) order}
We can further extend the unitary loop model, $U$, defined above to a (meta)-stable MBL phase~\footnote{MBL is expected to be only metastable in $2d$~\cite{de2017stability} due to an avalanche instability to rare thermalizing regions. However, at strong disorder, the time-scale for this avalanche stability is double-exponentially-long in disorder strength, and we will treat this ultra-long time scale as effectively infinite for practical purposes.} by flashing on an MBL Hamiltonian with eigenstate topological order equivalent to that of $\rm ICS_R$ after the final step. Specifically, modify: $U_{3}\rightarrow e^{-iH^{R}_\text{MBL}} U_{3}$, where: 
\begin{align}
H^{R}_\text{MBL} &= H_0+V \nonumber\\
H_0 &= -\sum_{\hexagon_p}\lambda_p \Phi_p 
- \sum_{ij\in R} \frac{1+\Phi_{p_{ij}}}{2} \mu_{ij} P_{ij}+\dots
\end{align}
where $\lambda,\mu$ are coupling constants with strong spatial randomness, and $``\dots"$ indicate generic perturbations that respect the emergent dynamical $e\leftrightarrow m$ symmetry of the bulk dynamics and  are much smaller than the typical size of $\lambda,\mu$. The flux factors in the last term have been chosen such that $[U,H_0]=0$. In the high-frequency limit (where the terms $H_0$ and $\lambda$ have small coefficients $\ll 1$), quasi-energy spectrum of the unitary evolution is given by the effective Hamiltonian $H_\text{eff}\approx H_0+V^S+\dots$, where $V^S = \frac12(V+U^\dagger V U)$ is the symmetrization of $V$ with respect to an (emergent-dynamical) $e\leftrightarrow m$ permutation symmetry, and 
$\dots$ denote higher-order corrections in the high-frequency expansion. Depending on the perturbations $\dots$, there are two possible fates for this model. The $e\leftrightarrow m$ symmetry could be spontaneously broken resulting in an MBL anyon time-crystal~\cite{Po_2017}, or resonances between the degenerate hybrid e/m excitations could lead to a breakdown of MBL. For details, we refer the reader to~\cite{Po_2017} which analyzes a similar MBL Hamiltonian for a topologically-equivalent Floquet Honeycomb model.

\section{Relating Floquet Codes and Continuous Loops}
For the HFC example above, we saw that there were continuously parameterized adiabatic or unitary loop(s) that i) pass through the same sequence of ICS's, and ii) implement the same operation  ($e\leftrightarrow m$ exchange) on the logical subspace of the code as for the measurement-only HFC model. 
This motivates us to define an equivalence relation between Floquet codes and loops based on if they all satisfy i) and ii).
In this section, we discuss some general considerations about whether and under what conditions an equivalence between FCs and loops might exist. Then, in the following section we formulate a specific class of FCs that generalize the HFC, for which we can directly establish an equivalence.

\subsection{Refining the notion of equivalence}
Before embarking on this we highlight two subtleties in defining an equivalence between codes and loops that arise in the HFC example. In the following, we will restrict our attention to local codes and loops, which can be generated by strictly local measurements or Hamiltonians respectively.

First, note that the trajectories of individual states generally differ between the measurement-only or parameterized unitary dynamics, and only the evolution of the logical subspace of the codes match. For example, the local measurement outcomes in each step of the HFC are random, so that, after one Floquet cycle of measurements a state initially in $\ICS_R$ returns to a state in $\ICS_R$ but with a generically-different pattern of fermion excitations on the red bonds. In contrast, the unitary dynamics produce deterministic state evolutions.
For this reason, we define criterion ii) above only in terms of the logical subspace.

Second, the correspondence between the HFC and ULs is one-to-many: there are multiple ULs that reproduce i) and ii) of the HFC, which differ by stacking with a non-topologically-ordered Floquet topological phase with rational CF index. 
The UL for this rational CF phase satisfies $U(t=1)=\mathbbm{1}$, implying that the eigenstates of $U(t=1)$ can be chosen to be short-range entangled. We will refer to ULs with this property as invertible, generalizing the terminology for ``integer" topological phases of gapped ground-states.
Here, by ``stacking", we mean that one can extend a unitary loop, $U(t+n)= U(t)$ by adding additional degrees of freedom and/or adding additional ``trivial" dynamics within the period. Specifically, we allow modifying the generating Hamiltonian, $H_0(t+1)=H_0(t)$ by adding an extra step:
\begin{align}
H_0(t)\rightarrow H(t)=\begin{cases}
2H_0(2t) & 0\leq t<1/2 \\
2H_1(2t) & t/2\leq t<1
\end{cases}
\end{align}
where $H_1(t)$ generates a trivial unitary loop $U_{H_1}\equiv \mathcal{T}e^{-i\int_0^1 H_1(t)dt}=\mathbbm{1}$. To get some intuition for this expression, note that the evolution for one period is: $U_{H} = U_{H_1}U_{H_0} = U_{H_0}$.
With this in mind, we define an equivalence class of ULs by moding out stacking with invertible Floquet phases, and look for a stable equivalence between FCs and these equivalence classes of ULs. 

\subsection{General considerations}
In general, Floquet codes and loops are \emph{not} equivalent in the sense defined above.

\paragraph{Chiral Topological Order}
For instance, ALs may have ground states with chiral topological orders (i.e. which have chiral edge states on an open manifold), which cannot be reached by measuring a sequence of local operators~\cite{Dubail_2015}. For this reason, we will exclude consideration of ICSs with chiral topological order.

\paragraph{Measurements can generate LRE} Another complication is that whereas loops generated by local Hamiltonians cannot modify the long-range entanglement (LRE) structure of a state, measurement-based dynamics can generate LRE. For instance, whereas a constant-depth unitary circuit cannot alter topological order, a constant depth measurement-circuit can convert between certain classes of short-range entangled and long-range entangled states~\cite{Kitaev_2006,tantivasadakarn2021long, verresen2021efficiently}. Hence, a local FC may exhibit a sequence of ICS's with distinct types of topological order, that cannot be traversed by constant depth, local adiabatic or unitary evolution. Several $3d$ examples  of FCs posses this property ~\cite{zhang2022x, Arpit2023} including one that hops between fracton ordered and conventional topological ordered states. \newline \indent
Consequently, to look for possible equivalences, we will demand that all ICSs of the FC have the same stably-equivalent topological order (i.e. equivalent up to adding short-range entangled degrees of freedom and entangling with a constant-depth circuit). \newline \indent
Having a homogeneous topology for the sequence of ICS's is still not yet sufficient to equate codes and loops. For example, given a loop that passes through a sequence of stabilizer-state ICS's, one may be tempted to define a Floquet code simply by measuring the stabilizers of each ICS. Yet, even with local measurements, one can end up effectively measuring non-local logical operators, thereby collapsing the encoded quantum information and ruining the code. For example, measuring the $x$-, $y$-, and $z$- bonds of the HFC rather than the $R$-, $G$-, and $B$- bonds results in the measurement of logical operators of the code. Hence, we will need to place additional constraints on the ICS to obtain an equivalence.
\paragraph{Relation between Adiabatic and Unitary Loops}
A UL, $U(t)$ always defines an AL: choose a base-point for the AL by a local, gapped Hamiltonian, $H(0)$ that commutes $U(t=1)$, and then define $H(\theta) = U^\dagger(2\pi t)H(0)U(2\pi t)$. 
For instance, when $U(1)$ implements an automorphism of a topological order, we may take $H$ to be a local Hamiltonian that realizes that topological order. However, the reverse is not obviously true: since an AL define the time evolution only on a single state (the ground-state), whereas defining a UL requires one to lift that action consistently to the entire spectrum of excited states. A sufficient condition for being able to lift an AL to a UL is if the AL is many-body localizable~\cite{Harper_2020}. Many-body localization (MBL) is not compatible with certain ingredients such as chiral or non-Abelian topological orders, continuous non-Abelian symmetries, or spontaneously broken continuous symmetries~\cite{Potter_2016}. Hence, in the following we will restrict our attention to systems with Abelian topological order, and with only Abelian symmetries (in fact, we will generally ignore symmetry throughout).

\section{Gauged, Paired Majorana Networks}
We next introduce a class of systems called paired defect networks, that generalize the structure of Kitaev's honeycomb model. Namely, their dynamics will implement an arbitrary anyon automorphism on the logical operators and local excitations.
We then establish a general equivalence between FCs, and ALs, ULs, and MBL FETs that pass through the same ICS's.
As explained above, to establish this equivalence, we restrict our attention to Abelian, non-chiral topological orders.

\subsection{Gauged fermion codes}
The Kitaev honeycomb model can be described as a gauged fermion system with Majorana ``defects" $c_i$ on each site $i$ of the honeycomb, and $\Z_2$ gauge connections $u_{ij}$ on the edges $\<ij\>$ of the honeycomb.

The Floquet honeycomb code has a particularly simple structure in this representation: the persistent stabilizers are the gauge flux through the plaquette, $\prod_{\<ij\> \in \hexagon} u_{ij}$, and the instantaneous check are projectors onto parity of Majorana pairs: $P_{ij}=ic_i u_{ij}c_j$.
After one cycle of measurements, the fluxes become frozen into a fixed non-dynamical pattern, and subsequently only the fermion degrees of freedom have non-trivial dynamics. In this case, we can regard $u_{ij}$ as a background, non-dynamical gauge field and consider just the system of Majorana fermions.

This structure can be extended to a network of Majoranas on vertices of a general graph (with even number of sites) with gauge links on edges. We define a paired-Majorana network code on such a graph as one stabilized by the flux through each face of the graph, and the fermion parities $P_{ij}$ of a fixed pairing of the Majoranas. Without loss of generality, we can consider only nearest-neighbor pairings (possibly by adding extra edges to the graph). Similarly, define a paired-Majorana network Floquet code as an FC for which each ICS is a paired-Majorana network code. 

\begin{figure}
    \centering
\includegraphics[width=0.6 \columnwidth]{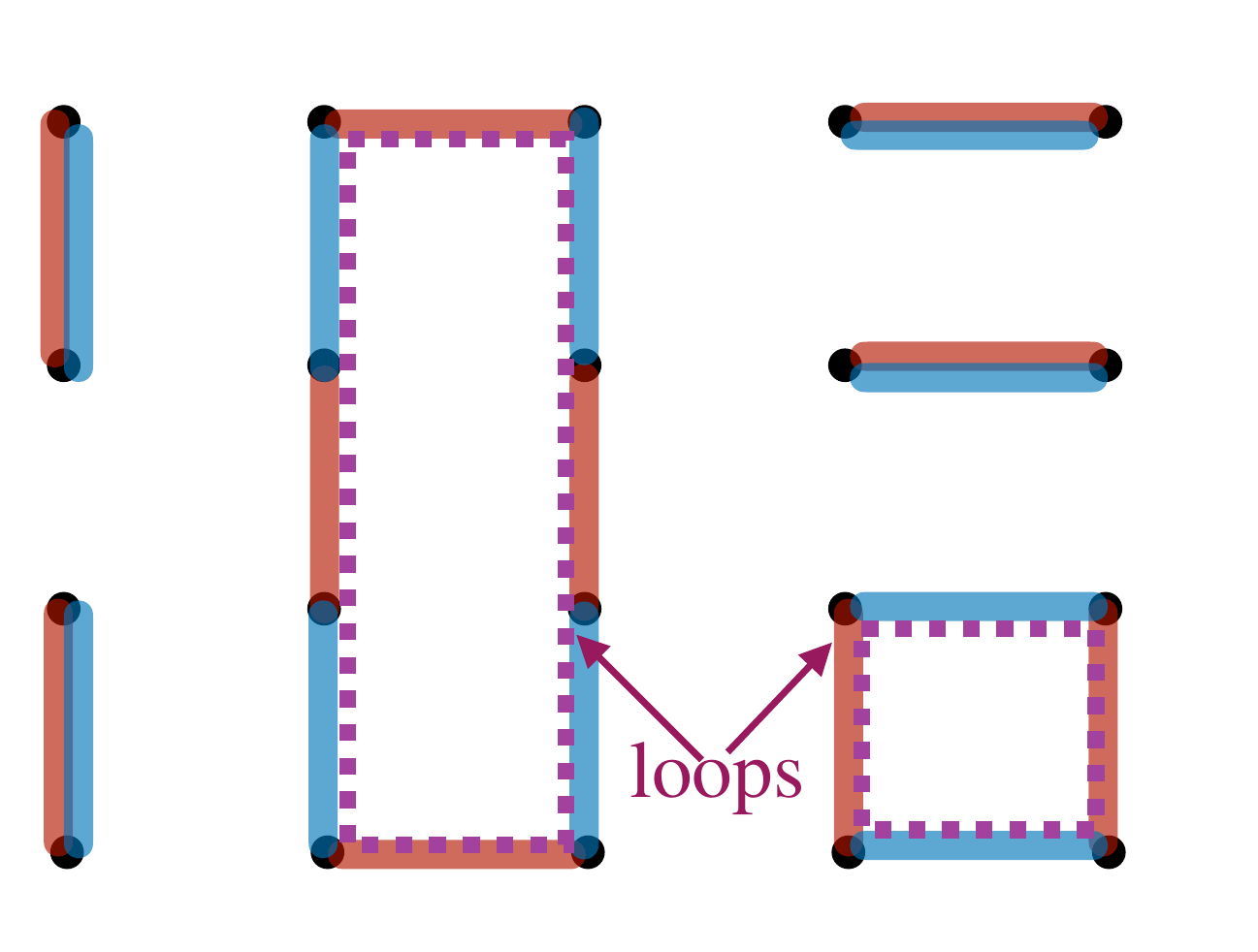}
    \caption{{\bf  Dimers and Loops -- } Each ICS of a paired Majorana network is represented as a dimer cover (shown here for a square lattice). Here, black dots represent Majoranas and red and blue bonds represent the pairings of two different ICS's. Transitions between two ICSs correspond to loop configurations (dashed lines). The appearance of non-contractible loops indicates that the logical operators can be inferred from the sequence of local dimer measurements.}
    \label{fig:dimersandloops}
\end{figure}

For this structure, there is a succinct condition for whether a given schedule of local Majorana-pair measurements results in the measurement of a logical operator. Each pairing defines a dimer cover of the graph, with each dimer representing an edge connecting the paired Majoranas. We can represent any dimer cover as a $\Z_2$ module on the edges of the graph, i.e. a formal sum of edges with $\Z_2$ coefficients. Consider the dimer covers $D_{i}$ and $D_{i+1}$ corresponding to subsequent ICSs in a FC. The sum $D_{i}+D_{i+1}$ defines a loop configuration on the graph. After subsequent measurement rounds, the gauge flux through each loop is also measured, since $\prod_{\<ij\>\in \text{loop}} P_{ij} = i^L\prod_{\<ij\>\in \text{loop}} u_{ij}\sim \Phi_\text{loop}$. Since logical operators of the code are given by gauge fluxes through non-contractible loops, the requirement that a measurement sequence does not measure a logical is that $D_i+D_{i+1}$ does not contain any non-contractable loops for any measurement round $i$. We refer to this as the ``no-long-loops" condition.

It is straightforward to establish an equivalence between Floquet codes, loops, and Floquet MBL phases for gauged Majorana defect networks satisfying this no-long-loops condition. 

\subsubsection{Adiabatic loops}
We begin by relating FCs and ALs. 
Given an AL that passes through ICSs that are paired Majorana network codes satisfying the no-long-loop condition, one can directly construct an FC that passes through this same sequence of ICSs simply by measuring the stabilizers for these ICSs.

Conversely, given a Majorana network FC satisfying the no-long-loops condition, we can define a gapped path between Hamiltonians $\{H_i\}$ with ground states in ICS$_i$ corresponding to pairing pattern $D_i$ with a fixed set of fermion parities for each bond with overall even fermion parity for all the bonds (in order to correspond to a valid state of the spin/qubit system). 
Namely, we can always consider interpolating from $H_i$ to $H_{i+1}$ separately within each loop of $D_i+D_{i+1}$ with no interactions between different loops.
This adiabatic path is gauged free-fermion Hamiltonian, which only has bilinear fermion interactions within the small loops of $D_i+D_{i+1}$. Hence, each loop has a finite size gap (except for accidental level crossings which can always be avoided) and the interpolation can be done adiabatically. The sequence of adiabatic paths defined by transitions between the ICSs therefore defines an AL that passes through the same sequence of ICSs. 

\subsubsection{Unitary loops}
To establish an equivalence to ULs, note that the change between $D_i$ and $D_{i+1}$ pairing patterns is equivalent to performing a permutation of the fermions within each loop of $D_i+D_{i+1}$, and applying a $\pm 1$ phase depending on the flux configuration. The no long loops condition implies that each of these loops are small, such that this can be performed with a finite depth unitary circuit, $U_{i,i+1}$. The sequence of unitaries defined by transitioning between the adjacent ICSs then define a local unitary evolution. 
This unitary evolution is not yet a loop: it preserves the dimer covering representing the first ICS, $D_1$, but, it may permute the different individual dimers making up $D_1$. 
We therefore seek to close the loop by adding additional steps of unitary evolution.

Denote the permutation of dimers in the first Floquet period as $g\in S_{N/2}$ where $N$ is the number of Majorana sites ($N$ is necessarily even). To undo the permutation, we can separately reverse each of the cycles of $g$ by adding additional dynamical steps that do not affect the logically-encoded information. Specifically, after a finite number of local unitary steps, each dimer can move at most a bounded distance away from its original position. Hence, each cycle is represented by a $1d$ loop through the dimers of $D_1$, where each loop segment has bounded size.
Therefore undoing each cycle of $g$ can be accomplished by a $1d$ quantum cellular automata (QCA)~\cite{Gross_2012} that executes a permutation of dimers. Such QCAs are exhaustively classified by a chiral index, and can always be written as the composition of local $1d$ Hamiltonian dynamics and chiral translation of dimers along the cycle.
Since the dimers are pairs of Majoranas, the chiral index takes rational values.
Moreover, all $1d$ QCAs acting on a closed region, $R$, can be generated by the boundary dynamics of a local $2d$ Hamiltonian acting on a $2d$ region $A$ that is bounded by $R$: $\d A=R$~\cite{Po_2016}  (note that we always consider a $2d$ or higher-dimensional system in order to have topological order).
Therefore, we can undo the permutation $g$, to close the sequence of unitaries in a loop, by stacking the system with rational CF phases on $2d$ subsystems. 
Since these rational CF phases do not affect the topological code space, they will preserve the dynamics within this logical space.

These considerations show that, for FC acting on paired Majorana networks, there are many unitaries that pass through the same ICSs and have the same dynamics within the logical space. However, these differ only by stacking with with $2d$ (invertible) rational CF phases.



\subsubsection{Floquet MBL phases}
Finally, since the ICSs of the paired Majorana codes are compatible with MBL, any UL acting on a paired Majorana network can be stabilized into an Floquet MBL phase by flashing on a disordered Hamiltonian consisting of the $H_{i} = \sum_{s \in S_i} \lambda_s S_i$ where $S_i$ are the (spatially local) stabilizers for the ICS for measurement round $i$, and $\lambda_s$ are spatially random couplings, to define the unitary evolution for one period: $U(t=1) = \prod_{i} U_{i,i+1}e^{-iH_i}$.

\begin{figure*}
        \centering
\includegraphics[width= 1.9\columnwidth]{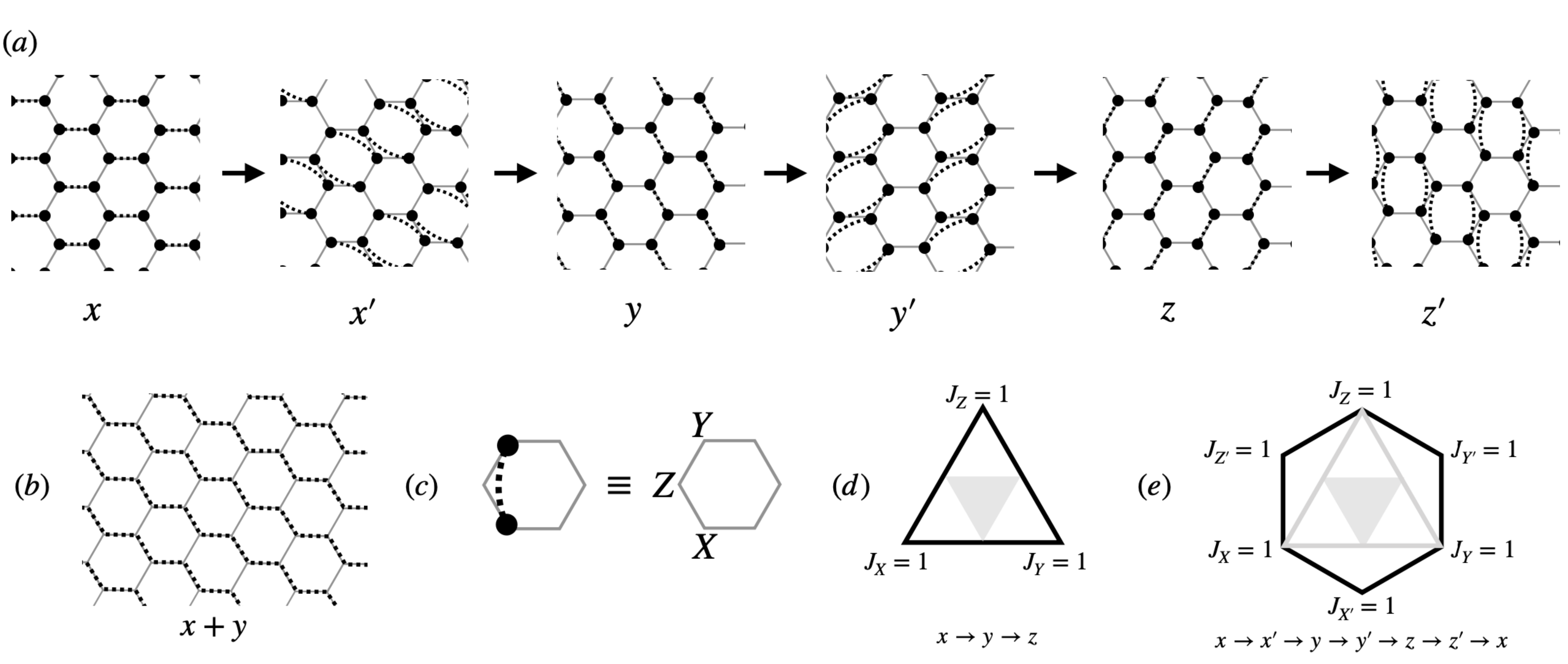}
    \caption{{\bf  Breaking long loops --} (a) The $x \to x^\prime \to y \to y^\prime \to z \to z^\prime$ schedule described in Sec \ref{sec:Breaking up long loops}. The ICSs correspond to toric code states and a full measurement cycle implements an $e-m$ swapping automorphism. (b) Long loops formed between $x$ and $y$ measurements in the $x \to y \to z$ schedule. (c) The primed rounds consist of next-nearest-neighbor parity measurements, which are three-body in the Pauli representation. We depict a measurement from round $z^\prime$ as an example. (d) The famous phase diagram of the Kitaev honeycomb model. Any closed loop interpolating between $J_X=1 \to J_Y =1 \to J_Z = 1 \to J_X =1$  must intersect the shaded grey region of gapless Hamiltonians. This implies the corresponding $x \to y \to z$ code will contain long loops. (e) By adding next-nearest-neighbor terms to the honeycomb model we can define an AL $J_X=1 \to J_{X^\prime} =1 \to J_Y = 1 \to J_{Y^\prime} =1 \to J_Z =1 \to J_{Z^\prime} \to J_X=1$, represented here by a hexagon which avoids the shaded gapless region. The existence of this AL indicates that the $x \to x^\prime \to y \to y^\prime \to z \to z^\prime$ code contains no long loops.}
    \label{fig:xyz_schedule}
\end{figure*}

\subsubsection{Breaking up long loops}
\label{sec:Breaking up long loops}
We next argue that, if a sequence of ICS's fails the no-long-loops condition, then it is possible to add intermediate ICS's that break up the long-loops with additional measurement rounds, possibly involving additional ancilla Majorana degrees of freedom freedom to the lattice that decouple from the original graph at each of the original ICSs.
Long loops arise when $D_i$ and $D_{i+1}$ differ by changing the $1d$ topological invariant of the Majoranas along a long loop $\ell$ (i.e. toggling $\ell$ between a topological and trivial superconducting wire).
We can break up this loop as follows. First, add ancilla copies of the Majorana fermions along the loop $\ell$, which are measured in the same pairing pattern as $D_i\cap \ell$ and $D_{i+1}\cap \ell$ in rounds $i$ and $i+1$ respectively. Since these ancilla do not pair with any of the original Majorana fermions at step $i$ and $i+1$, the modified ICSs for these measurment rounds differ simply by stacking a short-range entangled $1d$ fermion chain, and are hence stably-equivalent to the original ones. Then, to break up the long-loop formation, add an extra measurement round $i+1/2$, in which we pair each ancilla Majorana with its partner in the original system.
The modified sequence then satisfies the no-long-loops condition.

As an example, we illustrate this for a different implementation of the Honeycomb Floquet code, in which the bond-parities are measured on the $x\rightarrow y\rightarrow z\rightarrow x\rightarrow\dots$ bonds rather than on the Kekule-type $R\rightarrow G\rightarrow B\rightarrow R\rightarrow \dots$  bond measurement schedule.
Subsequently measuring $x$ and then $y$ bonds results in long diagonal loops (similarly for the transitions $y\rightarrow z$ and $z\rightarrow x$ between measurement rounds) as shown in Fig \ref{fig:xyz_schedule} (b). However, we may add additional measurement rounds which break up these loops with second nearest-neighbor measurements. The modified six round schedule $x \to x^\prime \to y \to y^\prime \to z \to z^\prime \to x \to \hdots$ is shown in Fig \ref{fig:xyz_schedule} a. In this case, no additional ancillas are needed; the primed rounds consist of fermion parity measurements between next nearest neighbor sites. In the qubit representation these next-to-nearest dimers correspond to three-body Pauli operators as can be seen in Fig \ref{fig:xyz_schedule} b. The Floquet code associated with this schedule evolves between six instantaneous code spaces, each of which can be mapped to the Wen-plaquette model \cite{WenPlaquette} on a square superlattice. After one full round the an $e-m$ exchanging automorphism is performed. This can be verified explicitly or through the edge dynamics of an associated unitary loop. The failure of the $s\to y\to z$ schedule to encode a logical subspace is related to the failure to satisfy the no-long-loops condition; the presence of the non-contractible loop in the ICS eliminates the non-local degeneracy. By equivalence discussed above this is further related the impossibility of an AL passing through the ICSs. This relationship is shown schematically the in Fig \ref{fig:xyz_schedule} c-d. By breaking up the long-loops the $x \to x^\prime \to y \to y^\prime \to z \to z^\prime \to x \to \hdots$ schedule has an associated AL which avoids the gapless points the $x\to y\to z$ loop is constrained to pass through.


\subsection{Application: topological index and dynamical anomalies for HFC boundaries}
\label{sec:HFC boundary}

\begin{figure*}
    \centering
\includegraphics[width= 1.8\columnwidth]{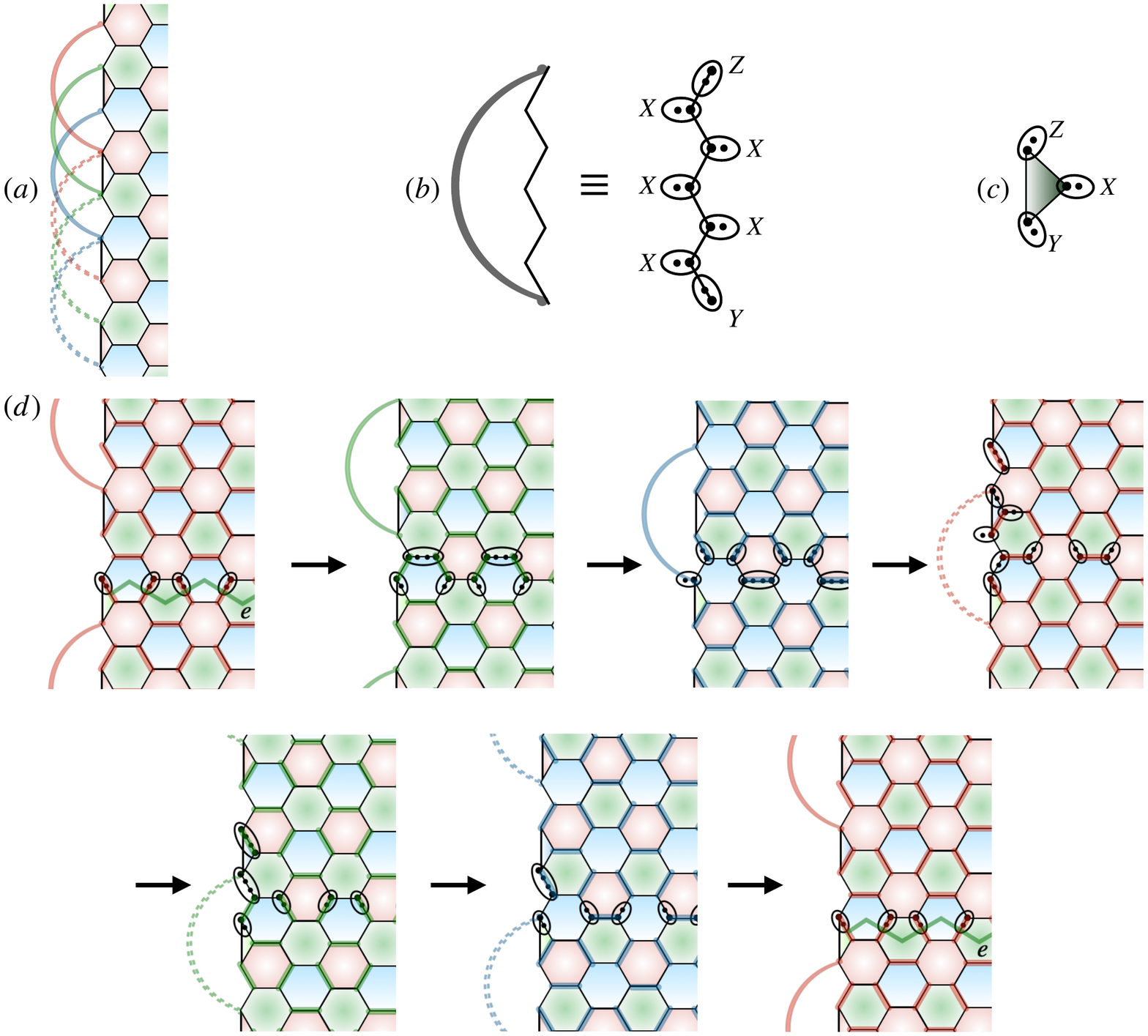}
    \caption{{\bf  Period-doubled gapped boundary of a planar HFC --} (a) HFC with boundary described in Sec. \ref{sec:HFC boundary}. Boundary measurement schedule has period six: solid red, solid green, solid blue, dashed red, dashed green, dashed blue, solid red... (b) Pauli string corresponding to the long arc checks on the boundary. (c) The flux through the triangles on the boundary are not inferred by the checks and must be explicitly measured. (d) Evolution of an $e$ logical operator terminating on the boundary. The check measurements performed in each round are highlighted. At each stage $r$ the logical operator is modified with the $r-$checks such that the resulting string commutes with the $r+1-$checks. After six rounds of evolution the operator remains an $e$ string. }
    \label{fig:hfcboundary}
\end{figure*}
As an application of this formalism, we next show how the equivalence between FCs and ULs places dynamical-anomaly constraints on the type of allowed boundaries of the qubit HFC.
Since any paired Majorana FC, such as the HFC or any modification that preserves this structure, is equivalent to a UL, they are both governed by the bulk-boundary correspondence and dynamical chiral-Floquet edge anomalies of the UL. Specifically, for ULs bulk $e\leftrightarrow m$ exchanging dynamics, implies a radical CF index $\nu(U) \in \sqrt{2}\mathbb{Q}$, which implies that there is no way to form a gapped/localized boundary~\cite{Po_2017} that preserves the time-translation symmetry of the bulk dynamics. The equivalence between FCs and ULs of paired-defect-network form, implies that there is similarly an obstruction to forming a gapped/logical edge of the HFC with a boundary measurement schedule of the same period as the bulk. 

Previous constructions for a planar HFC with gapped/logical boundary~\cite{Haah_Hastings_boundaries_2022} modified the bulk measurement sequence by doubling the periodicity and measuring $R\rightarrow G\rightarrow B\rightarrow G\rightarrow B\rightarrow R'$ bonds where $R$ and $R'$ differed at the edge. This sequence effectively performs the non-contractible loop $R\rightarrow G\rightarrow B$, and then undoes this loops by reversing its direction. The full 6-step sequence has no overall $e\leftrightarrow m$ exchanging action on logical operators. The unitarization of this process would be to consider alternating between chiral unitary evolution $U_1$, $\nu(U_1) = \sqrt{2}$ and antichiral unitary evolution, $U_2$ with $\nu_2 = 1/\sqrt{2}$, to get an overall trivial index, $\nu(U_2U_1)=1$ for the full evolution.

A related perspective comes from the equivalence of FCs and ALs for this class of systems. Namely, if we have an AL that exchanges $e\leftrightarrow m$ excitations, then there is no gapped boundary Hamiltonian with the same periodicity as the bulk. Specifically, there are two types of gapped boundaries of the $\Z_2$ gauge theory corresponding to condensing either $e$ or $m$ at the edge. Since one period of the HFC evolution exchanges $e$ and $m$ particles, it also exchanges these types of gapped boundaries. Hence, there is no periodic boundary-Hamiltonian that is both gapped and invariant under this $e\leftrightarrow m$ exchange. 

However, this argument suggests that it should be possible to gap the boundary by simply doubling the periodicity of the boundary without altering the bulk. Schematically, we could choose $H_\text{boundary}(t) = \begin{cases} H_e & 0\leq t<1 \\ H_m & 1\leq t < 2 \end{cases}$
where $H_{e,m}$ respectively represent boundary Hamiltonians for the $e,m$ condensed boundaries.
Returning to the FC setting, we can confirm that such a period-doubled gapped/logical boundary is indeed possible for the HFC. An example is drawn in Fig.~\ref{fig:hfcboundary}. Here, we consider a zig-zag edge of the Honeycomb, modified to a trivalent graph as shown in Fig.~\ref{fig:hfcboundary}(a). The check operators of each round are highlighted in Fig.~\ref{fig:hfcboundary}(d); the bulk measurements are unchanged ( $R \to G \to B$) and boundary measurement schedule has period six ( $\text{solid }R \to \text{solid }G \to \text{solid }B \to \text{dashed }R \to \text{dashed }G \to \text{dashed }B$). By inspection, one can see that each adjacent measurement round contains only short loops such that no logical operators are measured. By contrast, one can verify by inspection, that repeating the solid-line boundary conditions would result in the measurement of a long $f$-loop around the boundary upon going from the $B\rightarrow R$ step.

We note that, for this construction, one needs to directly measure gauge fluxes through the downward facing boundary triangle plaquettes shown in Fig. \ref{fig:hfcboundary}(c), as these are not accumulated as persistent stabilizers of the other measurements. Additionally, the boundary Majorana pair measurements are not nearest neighbor on the original lattice, and require several-spin measurements. These boundary measurements spoil the two-body measurement structure of the original HFC code. While potentially of practical importance for error correction thresholds, for the purposes of exploring general topological features of FCs, we view such details as non-universal engineering challenges.

One can also explicitly track the evolution of logical operators that terminate on the boundary as shown in Fig.~\ref{fig:hfcboundary}. The logical operators are either $e$ or $m$ strings that terminate on the open boundary. To form a logical qubit one needs to consider a plane with multiple holes punched out to form multiple non-contractible loops, however, for simplicity we simply show the end points of operators on one boundary. Starting with the solid-red ICS, we can label the logical operators as $e$ or $m$ strings depending on whether or not they terminate within or outside the red edge bond. To track the evolution of this operator to the next round, one needs to relabel it by tacking on $R$ stabilizers to form an operator that commutes with the $G$ measurements. Following the evolution through one bulk Floquet period, we see that the edge operator evolves as shown in Fig.~\ref{fig:hfcboundary}.

This example illustrates that, while the direct construction of gapped boundaries of the HFC can be complicated to construct, the insights from the ALs and ULs can identify possible universal mechanisms for their construction, rather than trying to build them by (potentially tedious) trial and error.

\section{General twist defects}
\label{sec: General twist defects}
We next aim to generalize the paired-Majorana structure of the HFC to other types of topological order.
To this end, a key step is to recognize the Majorana fermions in the Kitaev model as twist defects~\cite{Bombin2009fermions,Barkeshli_2013}, i.e. braiding an $e$ particle around a Majorana turns it into an $m$ particle and vice versa.
The notion of twist defects can be adapted to any Abelian topological order with anyon types $a,b,c\dots$, that posses an anyon automorphism $\sigma: a\rightarrow \sigma(a) $ which preserves the self- and mutual-statistics of the anyons: $\theta_{\sigma(a)}=\theta_a$, $\theta_{\sigma(a),\sigma(b)} = \theta_{a,b}$.
We refer to anyons that are uncharged by the automorphism (i.e $\sigma(a) = a$) as invariant anyons.

Numerous examples of such anyon automorphisms have been worked out for a large class of topological orders~\cite{Cheng_defects_2019}.
For our purposes, two representative examples are:

\begin{enumerate}
    \item {\it $\mathcal{D}_{\Z_N}$, with generalized $e\leftrightarrow m^q$ automorphism}:
    We denote a $\Z_N$ quantum double model, a.k.a. a $\Z_N$ toric code or $\Z_N$ gauge theory by $\mathcal{D}_{\Z_N}$. This theory has anyons: $\{e^jm^k\}$ with $j,k=0,1,2,\dots N-1$. There are automorphisms: $\sigma:e\mapsto m^q, m\mapsto e^p$ with $pq=1~\mod N$. Here, $d_\sigma=\sqrt{N}$ and the invariant anyons are $em^q$ or multiples thereof.
    
    \item {\it $\mathcal{D}_{\Z_N}^3$, with $S_3$ permutation automorphism}: $\Z_N^3$ gauge theory, i.e. 3-copies of $\Z_N$ toric code, with particles $\{1,e_i,m_i\dots\}$ with $i=1,2,3$ labeling the copy. This theory has an $S_3$ automorphism that permutes the different copies of the $\Z_N$ toric code. For example, we can consider twist defects that cyclically permute the copies: $\sigma: a_i\mapsto a_{i+1}$ for $a\in e,m,f$.
    This defect has quantum dimension $d_\sigma =  N^{4/3}$, and the bound state of like anyons in each copy (of the form $a_1a_2a_3$) are invariant.
\end{enumerate}
These twist defects are not deconfined excitations of the topological order, but rather are confined defects that must be written into the Hamiltonian. For example, they may occur at the ends of 1d ``wires" inside the topological phase. Further, twist defects always come in defect/anti-defect pairs. For example, in the $\Z_2$ gauge theory, Majorana twist defects can arise at the ends of a segment of topological superconducting wire made from the emergent fermions ($f$). 
Even though the underlying topological order is Abelian, the twist defects are non-Abelian~\cite{Bombin2009fermions, Cheng_defects_2019}, and have quantum dimension $d_\sigma >1$.

Define a defect network model to consist of a background topological order, with twist defects sitting at the vertices of some graph. A $2d$ defect network can be equivalently viewed as ``gluing" together open, simply-connected $2d$ patches (``cells") of topological order $a$ in patch $p$ to an topologically-equivalent anyon, $\phi_{p',p}(a)$ specified by some transition function $\phi$. A twist defect arises at the triple intersection of patches $p_1$, $p_1$, $p_3$ if $v_{p_3,p_2,p_1} = \phi_{p_1,p_3}\circ \phi_{p_3,p_2}\circ \phi_{p_2,p_1} \neq 1$. 
This cellular construction of defect networks was introduced by~\cite{else2019crystalline} to classify crystalline symmetry protected- and enriched- topological orders.  Here we adapt this approach to explore Floquet codes and phases. In the language of~\cite{else2019crystalline}, we will consider only invertible, point-like defects.
We note that the transition functions $\phi_{p',p}$ have a ``gauge" freedom under relabeling the anyons in $p$ and $p'$, however, the twist defect indicator, $v$ is invariant under such gauge transformations. This continuum approach will be explored in Sec. \ref{sec:continuum defect network}.

Defects can also be used to create a kind of parton construction to describe lattice models with local interactions following the ``slave-genon" approach of~\cite{Barkeshli_2015}.
This generalizes Kitaev's Majorana parton construction for the Honeycomb model to general twist defects~\footnote{Restricting to two body interactions this construction can only realize tripartite graphs, but a general local graph can be constructed by considering interactions between higher numbers of spins.}. In general, the local Hilbert space for these models will not be qubits, but rather qudits with $d_\sigma^2$ levels where $d_\sigma$ is the quantum dimension of the twist defect. We make use of this approach in Sec. \ref{sec:parton defect network} to realize lattice models for generalized FCs.  

In general we can define paired defect network codes as ones in which each twist defect $\sigma_i$ is paired with an anti-defect $\bar\sigma_j$ such that there is a definite fusion channel for the pair. Paired defect network FCs are then defined as those whose ICSs each have this property.
Note that the fusion product can always be measured by local operators using anyon interferometry. Namely, by creating $a,\bar{a}$ pairs, braiding the $a$ around the defect pair, and re-annihilating it with $\bar{a}$, and measuring the accumulated phase. These operations are local, commute with one another, and there always exists a set of anyons $a$ whose braiding phases uniquely determines the fusion outcome.

Following the same arguments presented for the Majorana-defect networks, for paired twist defect networks, there is an equivalence between FCs, ALs, ULs, and MBL FETs (note that the twist defect networks inside a background Abelian topological order are MBL-able since the fusion outcomes of these non-Abelian defects are always Abelian~\cite{Potter_2016}).

\subsection{Generalized Honeycomb Floquet codes from continuum twist defect networks}
\label{sec:continuum defect network}
\begin{figure*}
    \centering
    \includegraphics[width = 2\columnwidth]{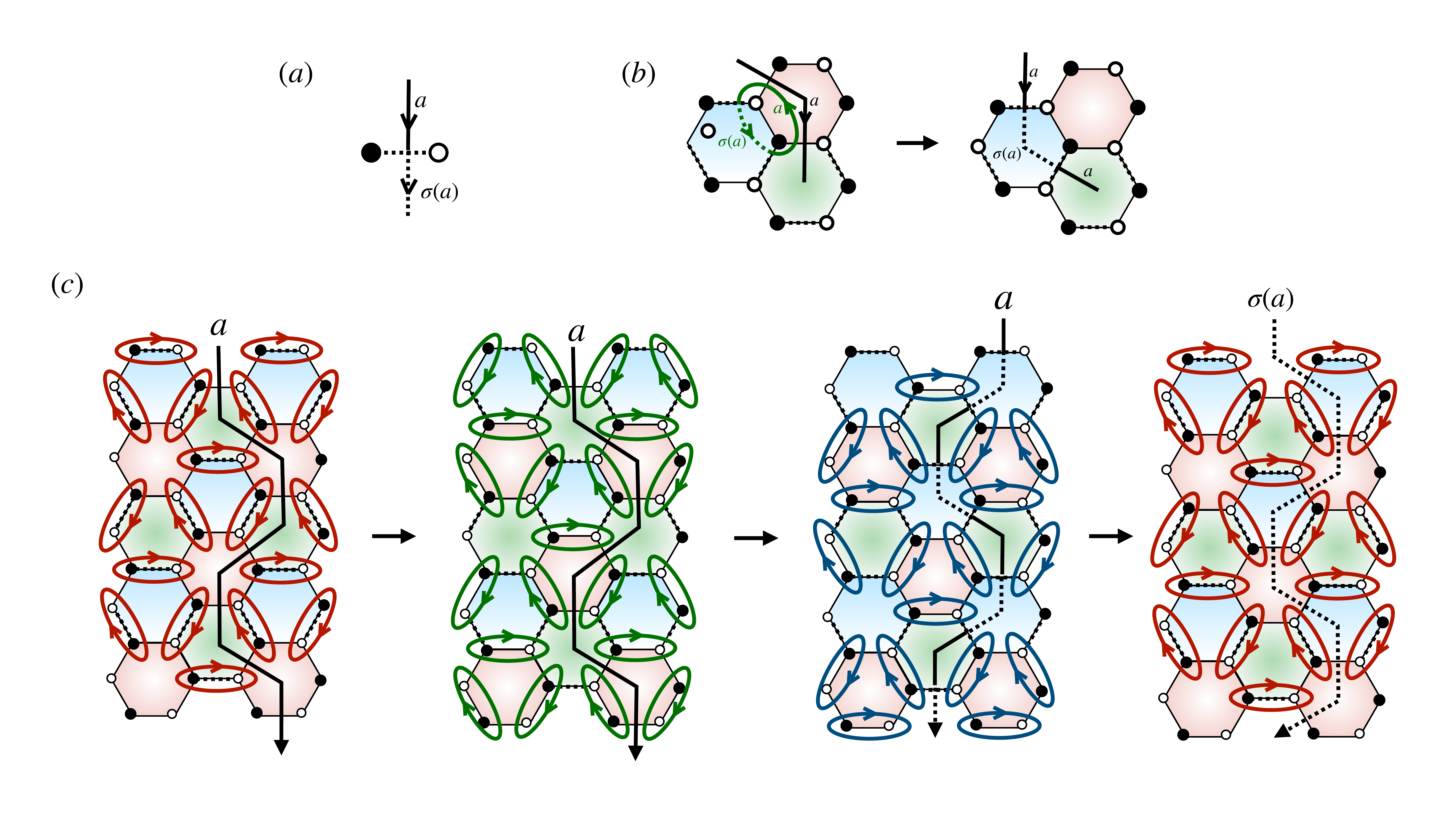}
    \caption{{\bf Evolution of logical operators in the generalized twist defect HFC -- } 
    (a). As anyon $a$ passes through edge connecting $\sigma$ (filled circle) and $\Bar{\sigma}$ (open circle) it is transformed into $\sigma(a)$. 
    (b). An anyon $a$ string can be moved through a pair of defect lines by multiplying it by a closed $\bar{a}$ loop which passes through the same defect lines. 
    (c). The colored loop around each bond represents the collection of all the braiding measurements carried out in each round. Since these have been measured, after each round one can employ the move shown in (b) to modify an anyon string so that it commutes with the measurements of the upcoming round. The net result is that the implementation of the automorphism $\sigma$ after one full cycle.
    }
    \label{fig:defect network}
\end{figure*}

To illustrate the construction of FCs and phases from twist-defect networks, we next construct a generalization of the Honeycomb code of Haah and Hastings to arbitrary non-chiral, Abelian topological order with non-trivial anyon permutation symmetry $\sigma$.~\footnote{We note that, at an abstract level, this construction can also describe adiabatic loops and Floquet codes with chiral or non-Abelian topological order.}

We start by forming a honeycomb by gluing together hexagonal plaquettes of a given topological order with twist defects $\sigma$ on each of the $A$ sublattice sites and anti-defects $\bar\sigma$ on each of the $B$ sublattice sites~\footnote{Here, we use $\sigma$ to denote both the defect and its associated anyon automorphism.}.
We then three-color the plaquettes with labels $R,G,B$, as was done with the qubit HFC. 
We choose the branch-cuts for the twist defects to reside along the red links, meaning that an anyons' topological charge gets transformed from $a$ to $\sigma(a)$ upon crossing a red link (with orientation shown in Fig.~\ref{fig:defect network}(a).


The measurement schedule follows that of the qubit HFC, measuring $R$ then $G$ then $B$ bonds. The Majorana parity measurements of Sec \ref{sec:Review of the HFC} are replaced with ``braiding check" measurements: we measure the phase that results from creating an anyon/anti-anyon pair, wrapping the anyon around a small loop enclosing the bond and then re-annihilating it with the anti-anyon. This anyon interferometry measurement partially determines the fusion channel of the $\sigma,\bar\sigma$ pair on each bond. In order to completely determine the fusion channel it is sufficient but not necessary (see Appendix \ref{app: GHFC measurement scheme}) to measure the aforementioned braiding process for each generating anyon $\{a_1,a_2,...a_N\}$ of the TO. In each round $r = R, G \text{ or }B$ a minimal set of braiding checks are performed around each bond of type $r$, collapsing the degeneracy introduced by the presence of the defects, and leaving behind a state of the ICS with only the topological degeneracy due to the underlying Abelian TO.


After each measurement round, logical operators of the ICS are anyon string operators that wrap around non-contractible loops which do not intersect any defect lines. They will need to be modified constantly by measured braiding checks in order to commute with the next round of measurements.  The evolution of logicals is depicted in Fig.~\ref{fig:defect network}, it is clear that after a full cycle of measurement an $a$ anyon string is converted into a $\sigma(a)$ string.


\subsubsection*{Persistent stabilizers and error correction}

 We can think of the patchwork defect model as a kind of topological quantum memory, where some quantum state is stored non-locally. From this point of view, open anyon strings correspond to errors: if one of these strings is allowed to wrap a non-contractible cycle a logical operation is performed, toggling the state of the system without our knowledge. To correct these errors we must be able to locally detect the presence of an open anyon string. 
 It follows that error correction requires some set of braiding measurements on the hexagonal patches of the defect network.
 
One could measure the phase obtained by braiding a complete set of generating anyons around each hexagonal plaquette. This proves to be excessive though, as some of these measurement outcomes can be inferred from persistent stabilizers which form from the braiding check measurements. These persistent stabilizers commute with every braiding check measurement, playing the role of the flux operators (see Fig. \ref{fig:HFC}) in the Majorana example.



\begin{figure}
    \includegraphics[width = 1\columnwidth]{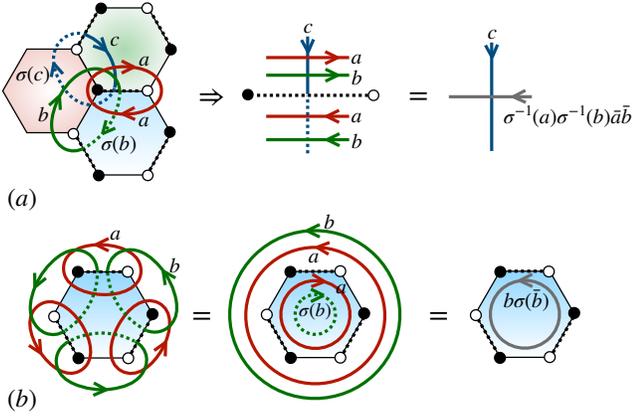}
    \caption{{\bf Persistent stabilizers --} (a). The blue check  represented as an anyon $c$ loop intersects the plaquette operator $P_{a,b}$ at 4 points, by deforming the anyon $a,b$ lines across the defect we obtain the effective anyon line of type $\sigma^{-1}(a)\sigma^{-1}(b)\bar{a}\bar{b}$ in gray, which must fuse to vacuum for it to commute with all the  blue checks: $\sigma^{-1}(a)\sigma^{-1}(b)\bar{a}\bar{b}=1\Rightarrow \sigma(ab)=ab.$  (b). The reduction of the persistent stabilizer $P_{a,b}$, notice the anyon $ab$ is invariant by the persistent stabilizer condition and can be moved inside the hexagon.}
    \label{fig:persistent stabilizer}
\end{figure}

Let us consider the exact form of these persistent stabilizers.  Imagine we have just completed the round $1$ and the braiding of anyon $a$ is measured on every red bond. Next, suppose anyon $b$ braiding is measured on every green bond in round $2$. At this point, plaquette stabilizers that are products of checks around any blue plaquette will be generated. We denote these plaquette operators by $P_{a,b}$. For $P_{a,b}$ to be persistent stabilizers it must commute with all subsequent measurements. In Fig \ref{fig:persistent stabilizer} a diagrammatic calculation is shown and the persistent stabilizer condition is found to be $\sigma(ab)=ab$. Using this condition the persistent stabilizer can be reduced to a $b\sigma(\bar{b})$ loop inside the plaquette. 

The braiding statistics between $b\sigma(\overline{b})$ and another anyon $c$ are given by
\begin{align}
    \theta_{c,b\sigma(\overline{b})}&=\theta_{c,b}-\theta_{c,\sigma(b)}\\
    &=\theta_{c,b}-\theta_{\sigma^{-1}(c),b}=\theta_{c\sigma^{-1}(\overline{c}),b}.
\end{align}
Therefore $c$ will commute with all persistent stabilizer when $\theta_{c\sigma^{-1}(\overline{c}),b}=0$ for all $b$. Thus implies $c\sigma^{-1}(\overline{c})=1$ or $ \sigma(c)=c$. From this we can conclude that the persistent stabilizers cannot detect invariant anyons. 

In order to detect errors corresponding to invariant anyons we can add supplemental braiding measurements of contractible loops inside the plaquettes of the defect lattice.  
Generically, if the TO has generating anyons $\{a_1,a_2,\hdots a_N\}$, braiding some subset $\{a_1,a_2, \hdots, a_Q\}$, where $Q<N$\footnote{ If the invariant anyons are generated by $\{f_1, f_2,\hdots , f_Q\} $ we can pick $Q<N$ generating anyons $\{a_1,a_2, \hdots, a_Q\}$ such that the braiding matrix $(\Theta_{fa})_{ij} = \Theta_{f_ia_j}$ has rank $Q$.}, will accomplish the task. To reiterate, these braiding measurements commute with all of braiding checks and so do not require a round of their own: they can be freely included in any of the R,G or B rounds. The choice of how the inclusion is carried out presumably affects fault tolerance properties. Generically though the resulting schedule is capable of detecting any local error.

\subsection{Generalized Honeycomb Floquet code lattice models from twist-defect partons}
\label{sec:parton defect network}
The above continuum description of the defect networks can be converted into an exactly-solvable lattice model whenever the underlying topological order admits such a lattice model description. As a brute-force construction, one could consider a network of patches of string-net models~\cite{Levin_2005}). However, the resulting lattice model will generally be cumbersome and involve measurement of many-spin terms. 

In this section we present an alternative defect network construction which naturally furnishes a lattice model description. The construction is based on the twist-defect parton (a.k.a. ``slave-genon") approach introduced by~\cite{Barkeshli_2015}. This approach yields generalized HFC models with non-chiral Abelian topological order with order-two twist defects (i.e. for which twisting twice restores the anyons). For this family of generalized HFC models we derive a measurement schedule corresponding to two-site nearest-neighbor Pauli measurements and describe the automorphism generated by measurements. Unlike the continuum description the parton construction only implements anyon automorphisms with order two ($\sigma^2 =1$).

Schematically, the twist-defect parton construction follows that of Kitaev's Majorana representation of spins-1/2, but replaces the Majorana defects of the $\Z_2$ topological order, with arbitrary twist defects, $\sigma$, of a general abelian topological order, $\TO_0$.
Parton constructions describe a local Hilbert space as a projection from a larger auxiliary Hilbert space.
Following~\cite{Barkeshli_2015}, the auxiliary Hilbert space can be viewed as a small island of topological order, $\TO_0$, with two twist defect/anti-defect ($\sigma/\bar\sigma)$ pairs.
The physical Hilbert space is then obtained by projecting each island (site) into the sector with trivial total topological charge. This projection plays the role of the gauge-constraint $c_ib^x_ib^y_ib^z_i=1$ in the spin-1/2 Kitaev model, which forces the four Majorana defects to have overall trivial fermion parity, yielding a bosonic spin model.
Here, a notable distinction from the original spin-1/2 Kitaev honeycomb model arises:
though in Kitaev's Honeycomb model, paired phases of the Majorana defects also had the same type of $\Z_2$ topological order, the Abelian (paired-defect) phase of generalized Kitaev models may have a completely different type of topological order, $\TO$, distinct from $\TO_0$. For example, ~\cite{Barkeshli_2015} used islands of fractional quantum Hall (FQH) bilayers, with interlayer-genon defects as the twist-defect partons, to construct a  generalized Kitaev honeycomb model with $\Z_N$ topological order.

\begin{figure*}
    \includegraphics[width = 1.4\columnwidth]{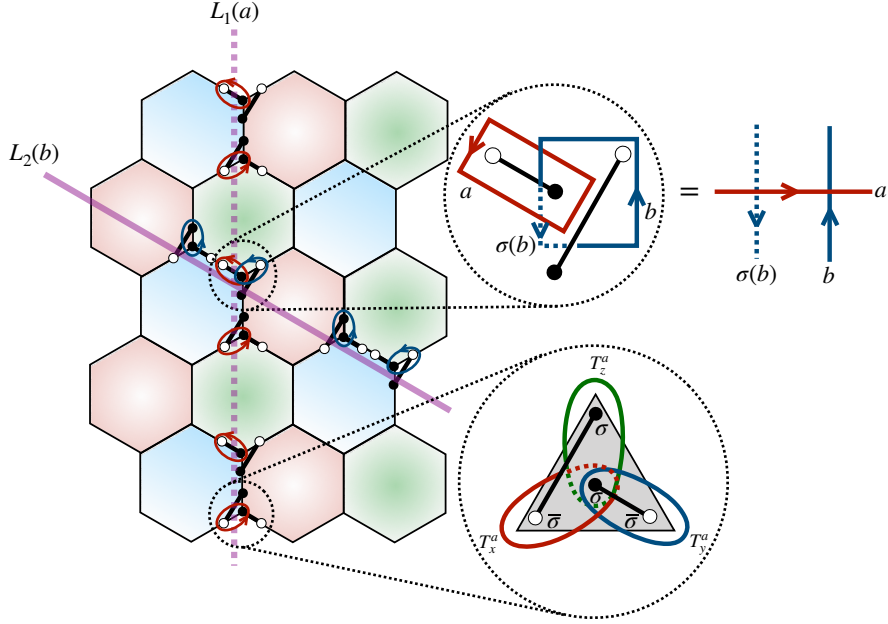}
    \caption{{\bf Generalized HFC --} The bottom right panel shows the micro structure of a site, which consists of 4 defects of $\TO_0$. The operators $T^a_{\alpha}$ are defined to be braiding of anyon $a$ around certain defect pair. Two logicals $L_1(a),L_2(a)$ after the red round measurement are depicted, with $L_1(a)$ being the product of red loops braiding along the path and $L_2(a)$ being the product of blue loops braiding along the path. More generally one defines an $L_1(a)$ string along a path that avoids the red plaquette, and whenever the string intersects a bond it takes half of the check operator on that bond. One defines an $L_2(a)$ string along a path that lies on the red bonds, and on each red bond $(ij)$ of type $\alpha_1$ one writes $T^{a}_{i,\alpha_2}T^{a}_{j,\alpha_3}$ where $\alpha_1,\alpha_2,\alpha_3$ are distinct. The two logicals intersect at a single site, which causes them fail to commute. The top right panel shows their commutation relation is: $L_1(a)L_2(b)=e^{2\pi i \theta_{a,b\sigma(\overline{b})}}L_2(b)L_1(a)$. It is straightforward to confirm from this construction that logicals of the same type commute and preserve fusion: $[L_i(a),L_i(b)]=0,~L_i(ab)=L_i(a)L_i(b)$. }
    \label{fig:lattice logicals}
\end{figure*}

In the parton description, the local, gauge-invariant ``spin" operators on each site: $T^a_{i,\alpha}$ are defined by braiding anyon $a$ around the pair $\alpha=x,y,z$ of defects on site/island $i$, as shown in Fig.~\ref{fig:lattice logicals}. 
The local operators obey the algebra:
\begin{align}
        &T^a_zT^b_x=e^{i\theta_{a,b\sigma(\overline{b})}}T^b_xT^a_z\\
        &T^a_xT^b_y=e^{i\theta_{a,b\sigma(\overline{b})}}T^b_xT^a_y\\
        &T^a_yT^b_z=e^{i\theta_{a,b\sigma(\overline{b})}}T^b_yT^a_z\\
        & T^a_\alpha T^b_\alpha=T^{ab}_\alpha,~ T^{a\dagger}_\alpha=T^{\overline{a}}_\alpha.
\end{align}
The gauge constraint $T^a_xT^a_yT^a_z=1$, implies that in the physical subspace $T^a_y\equiv T^{a\dagger}_xT^{a\dagger}_z$.
To form a local, on-site Hilbert space, we then construct a representation of this algebra for any abelian $\TO_0$ and $\sigma$. 

The anyons that are invariant under $\sigma$  form a subgroup of $\TO_0$\footnote{we use $\TO_0$ for both the abelian topological order itself and the group of its fusion rules, which is a finite abelian group.} that we denote by $\text{Inv}(\sigma):=\{a\in \TO_0,\sigma(a)=a\}$. Denote the quotient group $\TO_0/\text{Inv}(\sigma)$ by $\TO_0^\sigma$ and its elements, the equivalent class of $a$, by $[a]$. Let $\mH=\mathbb{C}[\TO_0^\sigma]$ be the group algebra over $\TO_0^\sigma$, that is, the formal linear combinations of group elements with complex coefficients. $\mH$ has a standard basis $\{~\left\vert[a]\right\rangle;~a\in\TO_0\}$ and has dimension $D=|\TO_0|/|\text{Inv}(\sigma)|$. Define $T^a_x,T^a_z$ by their action on the basis as:
\begin{align}
T^a_z|[g]\rangle=e^{i\theta_{a,\bar{g}\sigma(g)}}|[g]\rangle,~ T^a_x|[g]\rangle=|[ag\rangle]\rangle.
\label{eq:rep of local ops}
\end{align}
This representation is well-defined, since the phase factor of $T^a_z$ does not depend on the choice of $g$: $\theta_{a,\bar{g}\sigma(g)}$ is zero for any $g\in \text{Inv}(\sigma)$, and it forms the desired representation.

 Note that, in this representation, $T^a_\alpha=1$ for any $a\in \text{Inv}(\sigma)$. Namely: $T^a_z=1$ because the phase factor in Eq.~\ref{eq:rep of local ops} satisfies $\theta_{a,\overline{g}\sigma(g)}=\theta_{g,\overline{a}\sigma(a)}=0$,  and $T^a_x=1$ since $[ag]=[a][g]=1[g]=[g]$. In all examples known to us, the single site Hilbert space $\mH$ is a tensor product $\mathbb{C}_{d_1}\otimes \mathbb{C}_{d_2}\otimes\cdots$ and the operators $T^a_\alpha$ are products of generalized Paulis. For examples we refer the reader to appendix \ref{app: generalized HFC examples}.

\subsubsection{Measurement scheme and automorphisms}
The measurement schedule mimics that of HFC: plaquettes and bonds are 3-colored by $R,G,B$ labels. In each round check operators $T^a_{i,\alpha_{ij}}T^a_{j,\alpha_{ij}}$ on bonds of certain color are measured for all anyons $a$. 
We analyze the persistent stabilizers of this code in Appendix~\ref{app: generalized HFC gapped proof}, and show that, together with the checks in any given round, they completely fix the local excitations of the ICS, leaving only a finite-dimensional global topological logical space corresponding to a certain topological order $\TO$.
The persistent stabilizers can be deduced by considering combinations of checks that commute with the checks of all the $R,G,B$ rounds. For example, as a candidate persistent stabilizer on the $B$ plaquette take checks around its $R$ and $G$ edges that form its perimeter, with  $T^a_{i}T^a_j$ checks on type-$R$ edges and $T^b_iT^b_j$ checks on type-$G$ edges. Denote such plaquette stabilizer as $\mP_{a,b}$. It commutes with all the $R$ and $G$ checks. The diagrammatic calculation shown in Fig.~\ref{fig:persistent stabilizer} shows the condition for this plaquette operator to commute with all $B$ checks is: $\sigma(ab)=ab$, which has solution $b=\sigma(a)$. Any other solution would only differ from $\sigma(a)$ by an invariant anyon, which will not affect $\mP_{a,b}$. Thus we can denote the persistent stabilizer as $\mP_{[a]}$(recall $T^a_\alpha=1$ for any invariant $a$, thus $\mP_a$ only depends on the equivalent class $[a]$). For each plaquette we now have $|\TO_0|/|\text{Inv}(\sigma)$ persistent stabilizers.

We next examine the logical operators of this code.
For each anyon $a$, two distinct logical string operators, $L_1(a),L_2(a)$ can be defined, by analogy to the qubit HFC.
These are depicted in Fig.~\ref{fig:lattice logicals}, and satisfy 
commutation relations:
\begin{align}
    &[L_i(a),L_i(b)]=0, L_{i}(ab)=L_{i}(a)L_{i}(b)\nonumber\\
    &L_1(a)L_2(b)=e^{2\pi i \theta_{a,b\sigma(\bar{b})}}L_2(b)L_1(a).
    \label{eq:string alg}
\end{align} 

We emphasize that the topological order $\TO_0$ and twist defects $\sigma$ in the parton construction are completely auxiliary degrees of freedom, and are generally distinct from the induced topological order of the code, $\TO$. We are now in a position to deduce the structure of $\TO$.
$L_i(a)$ are string operators that create anyon strings of $\TO$. Denote the anyon of $\TO$ generated by the loop $L_i(a)$ with $a\in \TO_0$ as:  $\F_i(a)$. For Abelian topological orders, the fusion rules and mutual statistics, $\theta_{\F(a),\F(b)}$ can be determined purely from the algebra of the anyon string operators. In our case we have $\F_i(a)\times \F_i(b)=\F_i(ab)$ and the non-trivial mutual statistics are $\theta_{\F_1(a),\F_2(b)}=\theta_{a,\overline{b}\sigma(b)}$.  
Invariant anyons of $\{\TO_0,\sigma\}$ are mapped to vacuum: $\F_i(a)=1$ if $\sigma(a)=a$, since $\F_i(a)$ has trivial braiding with all other anyons. Therefore each type of string operator is capable of creating $|\TO_0|/|\text{Inv}(\sigma)|=D$ anyons, $\TO$ will have in total $|\TO|=D^2=(|\TO_0|/|\text{Inv}(\sigma)|)^2$ anyons (including the vacuum sector). 

Having worked out the induced topological order, $\TO$, we next examine the Floquet dynamics of logical operators (``logicals").
Following similar reasoning as in the HFC model one will find that that there is a permutation: $L_1(a)\leftrightarrow L_2(a)$ after each Floquet period. This corresponds to an order-two anyon permutation of $\TO$: $\F_1(a)\leftrightarrow \F_2(a)$, which we denote as $\varphi$. Moreover $\varphi$ is  an automorphism of $\TO$ since it preserves all anyon braiding statistics:
\begin{align}
    \theta_{\F_1(a),\F_2(b)}&=\theta_{a,\overline{b}\sigma(b)}=-\theta_{a,b}+\theta_{a,\sigma(b)}\\
    &=-\theta_{a,b}+\theta_{\sigma(a),b}=\theta_{\overline{a}\sigma(a),b}
\end{align}
which is equal to the exchanged value: $\theta_{\F_2(a),\F_1(b)}=\theta_{b,\overline{a}\sigma(a)}$.

In Appendix~\ref{app: generalized HFC examples} we derive a few different examples of the topological order $\TO,\varphi$, induced for an HFC with on-site degrees of freedom described by $\sigma$ twist-defects partons of auxiliary topological order $\TO_0$. The results are summarized in table \ref{table: lattice model}.
\begin{table}[h]
\centering
\setlength{\tabcolsep}{12pt} 
\renewcommand{\arraystretch}{1.3} 
\begin{tabular}{|m{3.5cm}|m{3.5cm}|}
\hline
{\footnotesize \bf Auxiliary topological order and twist defects partons on sites: $\TO_0,\sigma$ }
& 
{\footnotesize \bf Resulting topological order in paired-defect network ICS's and automorphisms: $\TO,\varphi$}  \\
\hline
$U(1)_N\times U(1)_N$, $a_1\leftrightarrow a_2$ & $\mathcal{D}_{\mathbb{Z}_N}, e\leftrightarrow m$ \\
\hline
$\mathcal{D}_{\mathbb{Z}_N},a\to \overline{a},N=2n>2$ & $\mathcal{D}_{\mathbb{Z}_n}^2, a_1\leftrightarrow a_2$ \\
\hline
$\mathcal{D}_{\mathbb{Z}_N},a\to \overline{a}, N=2n-1$& $\mathcal{D}_{\mathbb{Z}_N}^2$, $e_1\to e_2^2,m_1\to m_2^n, e_2\to e_1^n, m_2\to m_1^2$ \\
\hline
$\mathcal{D}_{\mathbb{Z}_N}, e\to m^p, m\to e^q, pq=1\mod N$ & $\mathcal{D}_{\mathbb{Z}_N}, e\to \overline{m}^p, m\to \overline{e}^q$\\
\hline
\end{tabular}
\caption{Notation: $U(1)_N$ is the $\frac{1}{N}$ Laughlin state, $\mathcal{D}_{\mathbb{Z}_N}$ stands for $\mathbb{Z}_N$ toric code, subscripts on anyons are layer labels.
}
\label{table: lattice model}
\end{table}

\subsubsection{Unitarization and chiral floquet index}
Just as for the spin-1/2 version, the measurement schedule of these generalized HFC models satisfies all the requirements of a fully-paired twist-defects with no-long loops in transitions between ICSs. Hence, by the general results above, we can lift this measurement-only FC into a unitary loop. In fact, we may directly follow the constructions of Section~\ref{sec: majorana lift}, replacing the spin-1/2 Pauli algebra by the generalized operators $T^a_{x,y,z}$.
Just as for the qubit HFC, 
this results in a ``radical" chiral FET~\cite{Po_2017} exhibiting chiral translation of $\varphi$ twist-defects around a spatial boundary~\footnote{A subtlety is that $\sigma$ is a twist-defect of the auxiliary parton construction topological order $\TO_0$, which generally not equal to the induced topological order, $\TO$ and twist defect $\varphi$ of the resulting FET.
However, it turns out that $d_\varphi = d_\phi$.
To see this note that, for $\sigma$: $\sigma\times \sigma=\sum_a a\sigma(\overline{a})$, which has $|\TO_0|/\text{Inv}(\sigma)$ distinct terms, therefore  $d_\sigma=\sqrt{|\TO_0|/\text{Inv}(\sigma)}$. For $\varphi: \F_1(a)\leftrightarrow \F_2(a)$, therefore $\varphi\times \varphi=\sum_{a}\F_1(\overline{a})\F_2(a)$, which also has $|\TO_0|/\text{Inv}(\sigma)$ different terms(recall $\F(a)=1$ for any $a\in \text{Inv}(\sigma)$).}. 
The corresponding chiral Floquet index was analyzed in~\cite{Po_2017}, and takes value $\chi(U)= d_\varphi\mathbb{Q}$. We note that while the chiral Floquet index has only been rigorously defined for trivial or $\Z_2$ topological orders, we expect that it can be generalized to arbitrary topological orders and anyon models. For the present models, with order-two twist defects (which have $d_\sigma$ that are square-root of an integer), we may sidestep this difficulty by considering $U^2$ which is an invertible CF order, with index $\nu(U^2) = \nu(U)^2 = d_\varphi^2 \mathbb{Q}^2$. 

This result supports the conjecture~\cite{Po_2017} that a Floquet MBL system realizing a bulk topological order automorphism will have edge chiral index whose irrational part is given by the quantum dimension of the corresponding defect. Similar to what we showed for HFC, this nontrivial radical CF index then put constraints on the possible boundary dynamics of the generalized HFC, namely a gapped boundary is only possible with doubled periodicity.  

\section{Discussion}
The defect network constructions introduced in this work provide a direct connection between Floquet codes and Floquet enriched topological orders. 
These results establish a throughline connecting topological indices for Floquet phases and practical issues for designing quantum error correcting codes. 

Our results suggest a number of avenues for further exploration:

While we have mainly focused on $2d$ models with Abelian topological order based on generalizations of the Kitaev Honeycomb model, it may be interesting to extend these constructions to non-Abelian systems capable of universal topological computation, or to $3d$ \cite{XcubeFloquet_2022,Arpit2023,Williamson2022spacetime} where the theory of twist-defects, topological order, and fracton-orders are less well characterized.

From a practical quantum error correction perspective, it would be desireable to develop simplified lattice models for general twist defect networks, and to design possible physical realizations of generalized Floquet codes in qubit arrays, AMO quantum simulators, or correlated electron materials. A second challenge is to understand the resulting code properties, such as the universality class and scaling properties of their error-correcting threshold phase transitions, and the practical error-correction thresholds for realistic implementations and decoders. Some progress \cite{Fisher2023} \cite{Gidney2021faulttolerant} has been made in these directions already.

One potential way of enhancing the quantum storage capacity of FCs would be to introduce defects into the TO state encoded in each of the ICSs. In the case of the HFC and its $\mathbb{Z}_N$ generalizations this has been worked out explicitly \cite{FCtwist_2023}. Here the measurement cycle produces a sequence of ICSs equivalent to toric code with lattice dislocations. It would be interesting to understand a prescription for introducing defects into ICSs of the general class of FCs discussed in this work. It seems somewhat natural that the defect network constructions we have considered should be well suited to this task.


\vspace{24pt} \noindent{\it Acknowledgements --} We thank David Aasen, Arpit Dua, Tyler Ellison, Nat Tantivasadakarn, and Dominic Williamson for insightful discussions. This work was supported by DOE DE-SC0022102 (JS,ACP), and in part by the Alfred P. Sloan Foundation through a Sloan Research Fellowship (ACP). This work was partly performed at KITP supported by the National Science Foundation under Grant No. NSF PHY-1748958.

\vspace{24pt} \noindent{\it Related work --} While completing this manuscript, we learned about two forthcoming related works connecting Floquet codes and the chiral Floquet unitary index~\cite{roberts2023geometric, Dave2023}.

\bibliography{bib}

\begin{thebibliography}{43}%
\makeatletter
\providecommand \@ifxundefined [1]{%
 \@ifx{#1\undefined}
}%
\providecommand \@ifnum [1]{%
 \ifnum #1\expandafter \@firstoftwo
 \else \expandafter \@secondoftwo
 \fi
}%
\providecommand \@ifx [1]{%
 \ifx #1\expandafter \@firstoftwo
 \else \expandafter \@secondoftwo
 \fi
}%
\providecommand \natexlab [1]{#1}%
\providecommand \enquote  [1]{``#1''}%
\providecommand \bibnamefont  [1]{#1}%
\providecommand \bibfnamefont [1]{#1}%
\providecommand \citenamefont [1]{#1}%
\providecommand \href@noop [0]{\@secondoftwo}%
\providecommand \href [0]{\begingroup \@sanitize@url \@href}%
\providecommand \@href[1]{\@@startlink{#1}\@@href}%
\providecommand \@@href[1]{\endgroup#1\@@endlink}%
\providecommand \@sanitize@url [0]{\catcode `\\12\catcode `\$12\catcode
  `\&12\catcode `\#12\catcode `\^12\catcode `\_12\catcode `\%12\relax}%
\providecommand \@@startlink[1]{}%
\providecommand \@@endlink[0]{}%
\providecommand \url  [0]{\begingroup\@sanitize@url \@url }%
\providecommand \@url [1]{\endgroup\@href {#1}{\urlprefix }}%
\providecommand \urlprefix  [0]{URL }%
\providecommand \Eprint [0]{\href }%
\providecommand \doibase [0]{https://doi.org/}%
\providecommand \selectlanguage [0]{\@gobble}%
\providecommand \bibinfo  [0]{\@secondoftwo}%
\providecommand \bibfield  [0]{\@secondoftwo}%
\providecommand \translation [1]{[#1]}%
\providecommand \BibitemOpen [0]{}%
\providecommand \bibitemStop [0]{}%
\providecommand \bibitemNoStop [0]{.\EOS\space}%
\providecommand \EOS [0]{\spacefactor3000\relax}%
\providecommand \BibitemShut  [1]{\csname bibitem#1\endcsname}%
\let\auto@bib@innerbib\@empty
\bibitem [{\citenamefont {Hastings}\ and\ \citenamefont
  {Haah}(2021)}]{HH_dynamic_2021}%
  \BibitemOpen
  \bibfield  {author} {\bibinfo {author} {\bibfnamefont {M.~B.}\ \bibnamefont
  {Hastings}}\ and\ \bibinfo {author} {\bibfnamefont {J.}~\bibnamefont
  {Haah}},\ }\bibfield  {title} {\bibinfo {title} {Dynamically generated
  logical qubits},\ }\href {https://doi.org/10.22331/q-2021-10-19-564}
  {\bibfield  {journal} {\bibinfo  {journal} {Quantum}\ }\textbf {\bibinfo
  {volume} {5}},\ \bibinfo {pages} {564} (\bibinfo {year} {2021})}\BibitemShut
  {NoStop}%
\bibitem [{\citenamefont {Paetznick}\ \emph {et~al.}(2023)\citenamefont
  {Paetznick}, \citenamefont {Knapp}, \citenamefont {Delfosse}, \citenamefont
  {Bauer}, \citenamefont {Haah}, \citenamefont {Hastings},\ and\ \citenamefont
  {da~Silva}}]{Paetznick_2023}%
  \BibitemOpen
  \bibfield  {author} {\bibinfo {author} {\bibfnamefont {A.}~\bibnamefont
  {Paetznick}}, \bibinfo {author} {\bibfnamefont {C.}~\bibnamefont {Knapp}},
  \bibinfo {author} {\bibfnamefont {N.}~\bibnamefont {Delfosse}}, \bibinfo
  {author} {\bibfnamefont {B.}~\bibnamefont {Bauer}}, \bibinfo {author}
  {\bibfnamefont {J.}~\bibnamefont {Haah}}, \bibinfo {author} {\bibfnamefont
  {M.~B.}\ \bibnamefont {Hastings}},\ and\ \bibinfo {author} {\bibfnamefont
  {M.~P.}\ \bibnamefont {da~Silva}},\ }\bibfield  {title} {\bibinfo {title}
  {Performance of planar floquet codes with majorana-based qubits},\ }\bibfield
   {journal} {\bibinfo  {journal} {{PRX} Quantum}\ }\textbf {\bibinfo {volume}
  {4}},\ \href {https://doi.org/10.1103/prxquantum.4.010310}
  {10.1103/prxquantum.4.010310} (\bibinfo {year} {2023})\BibitemShut {NoStop}%
\bibitem [{\citenamefont {Haah}\ and\ \citenamefont
  {Hastings}(2022)}]{Haah_Hastings_boundaries_2022}%
  \BibitemOpen
  \bibfield  {author} {\bibinfo {author} {\bibfnamefont {J.}~\bibnamefont
  {Haah}}\ and\ \bibinfo {author} {\bibfnamefont {M.~B.}\ \bibnamefont
  {Hastings}},\ }\bibfield  {title} {\bibinfo {title} {Boundaries for the
  honeycomb code},\ }\href {https://doi.org/10.22331/q-2022-04-21-693}
  {\bibfield  {journal} {\bibinfo  {journal} {Quantum}\ }\textbf {\bibinfo
  {volume} {6}},\ \bibinfo {pages} {693} (\bibinfo {year} {2022})}\BibitemShut
  {NoStop}%
\bibitem [{\citenamefont {Vuillot}(2021)}]{vuillot2021planar}%
  \BibitemOpen
  \bibfield  {author} {\bibinfo {author} {\bibfnamefont {C.}~\bibnamefont
  {Vuillot}},\ }\bibfield  {title} {\bibinfo {title} {Planar floquet codes},\
  }\href@noop {} {\bibfield  {journal} {\bibinfo  {journal} {arXiv preprint
  arXiv:2110.05348}\ } (\bibinfo {year} {2021})}\BibitemShut {NoStop}%
\bibitem [{\citenamefont {Aasen}\ \emph {et~al.}(2022)\citenamefont {Aasen},
  \citenamefont {Wang},\ and\ \citenamefont {Hastings}}]{Aasen2022adiabatic}%
  \BibitemOpen
  \bibfield  {author} {\bibinfo {author} {\bibfnamefont {D.}~\bibnamefont
  {Aasen}}, \bibinfo {author} {\bibfnamefont {Z.}~\bibnamefont {Wang}},\ and\
  \bibinfo {author} {\bibfnamefont {M.~B.}\ \bibnamefont {Hastings}},\
  }\bibfield  {title} {\bibinfo {title} {Adiabatic paths of hamiltonians,
  symmetries of topological order, and automorphism codes},\ }\href
  {https://doi.org/10.1103/PhysRevB.106.085122} {\bibfield  {journal} {\bibinfo
   {journal} {Phys. Rev. B}\ }\textbf {\bibinfo {volume} {106}},\ \bibinfo
  {pages} {085122} (\bibinfo {year} {2022})}\BibitemShut {NoStop}%
\bibitem [{\citenamefont {Kesselring}\ \emph {et~al.}(2022)\citenamefont
  {Kesselring}, \citenamefont {de~la Fuente}, \citenamefont {Thomsen},
  \citenamefont {Eisert}, \citenamefont {Bartlett},\ and\ \citenamefont
  {Brown}}]{brown2022}%
  \BibitemOpen
  \bibfield  {author} {\bibinfo {author} {\bibfnamefont {M.~S.}\ \bibnamefont
  {Kesselring}}, \bibinfo {author} {\bibfnamefont {J.~C.~M.}\ \bibnamefont
  {de~la Fuente}}, \bibinfo {author} {\bibfnamefont {F.}~\bibnamefont
  {Thomsen}}, \bibinfo {author} {\bibfnamefont {J.}~\bibnamefont {Eisert}},
  \bibinfo {author} {\bibfnamefont {S.~D.}\ \bibnamefont {Bartlett}},\ and\
  \bibinfo {author} {\bibfnamefont {B.~J.}\ \bibnamefont {Brown}},\ }\bibfield
  {title} {\bibinfo {title} {Anyon condensation and the color code}\ }\href
  {https://doi.org/10.48550/ARXIV.2212.00042} {10.48550/ARXIV.2212.00042}
  (\bibinfo {year} {2022})\BibitemShut {NoStop}%
\bibitem [{\citenamefont {Levin}\ and\ \citenamefont {Wen}(2005)}]{Levin_2005}%
  \BibitemOpen
  \bibfield  {author} {\bibinfo {author} {\bibfnamefont {M.~A.}\ \bibnamefont
  {Levin}}\ and\ \bibinfo {author} {\bibfnamefont {X.-G.}\ \bibnamefont
  {Wen}},\ }\bibfield  {title} {\bibinfo {title} {String-net
  condensation:{\hspace{1em} }a physical mechanism for topological phases},\
  }\bibfield  {journal} {\bibinfo  {journal} {Physical Review B}\ }\textbf
  {\bibinfo {volume} {71}},\ \href {https://doi.org/10.1103/physrevb.71.045110}
  {10.1103/physrevb.71.045110} (\bibinfo {year} {2005})\BibitemShut {NoStop}%
\bibitem [{\citenamefont {Kitaev}(2006)}]{Kitaev_2006}%
  \BibitemOpen
  \bibfield  {author} {\bibinfo {author} {\bibfnamefont {A.}~\bibnamefont
  {Kitaev}},\ }\bibfield  {title} {\bibinfo {title} {Anyons in an exactly
  solved model and beyond},\ }\href {https://doi.org/10.1016/j.aop.2005.10.005}
  {\bibfield  {journal} {\bibinfo  {journal} {Annals of Physics}\ }\textbf
  {\bibinfo {volume} {321}},\ \bibinfo {pages} {2} (\bibinfo {year}
  {2006})}\BibitemShut {NoStop}%
\bibitem [{\citenamefont {Kong}\ and\ \citenamefont
  {Wen}(2014)}]{kong2014braided}%
  \BibitemOpen
  \bibfield  {author} {\bibinfo {author} {\bibfnamefont {L.}~\bibnamefont
  {Kong}}\ and\ \bibinfo {author} {\bibfnamefont {X.-G.}\ \bibnamefont {Wen}},\
  }\bibfield  {title} {\bibinfo {title} {Braided fusion categories,
  gravitational anomalies, and the mathematical framework for topological
  orders in any dimensions},\ }\href@noop {} {\bibfield  {journal} {\bibinfo
  {journal} {arXiv preprint arXiv:1405.5858}\ } (\bibinfo {year}
  {2014})}\BibitemShut {NoStop}%
\bibitem [{\citenamefont {Thouless}(1983)}]{Thouless_pump}%
  \BibitemOpen
  \bibfield  {author} {\bibinfo {author} {\bibfnamefont {D.~J.}\ \bibnamefont
  {Thouless}},\ }\bibfield  {title} {\bibinfo {title} {Quantization of particle
  transport},\ }\href {https://doi.org/10.1103/PhysRevB.27.6083} {\bibfield
  {journal} {\bibinfo  {journal} {Phys. Rev. B}\ }\textbf {\bibinfo {volume}
  {27}},\ \bibinfo {pages} {6083} (\bibinfo {year} {1983})}\BibitemShut
  {NoStop}%
\bibitem [{\citenamefont {Gross}\ \emph {et~al.}(2012)\citenamefont {Gross},
  \citenamefont {Nesme}, \citenamefont {Vogts},\ and\ \citenamefont
  {Werner}}]{Gross_2012}%
  \BibitemOpen
  \bibfield  {author} {\bibinfo {author} {\bibfnamefont {D.}~\bibnamefont
  {Gross}}, \bibinfo {author} {\bibfnamefont {V.}~\bibnamefont {Nesme}},
  \bibinfo {author} {\bibfnamefont {H.}~\bibnamefont {Vogts}},\ and\ \bibinfo
  {author} {\bibfnamefont {R.~F.}\ \bibnamefont {Werner}},\ }\bibfield  {title}
  {\bibinfo {title} {Index theory of one dimensional quantum walks and cellular
  automata},\ }\href {https://doi.org/10.1007/s00220-012-1423-1} {\bibfield
  {journal} {\bibinfo  {journal} {Communications in Mathematical Physics}\
  }\textbf {\bibinfo {volume} {310}},\ \bibinfo {pages} {419} (\bibinfo {year}
  {2012})}\BibitemShut {NoStop}%
\bibitem [{\citenamefont {Rudner}\ \emph {et~al.}(2013)\citenamefont {Rudner},
  \citenamefont {Lindner}, \citenamefont {Berg},\ and\ \citenamefont
  {Levin}}]{rudner2013anomalous}%
  \BibitemOpen
  \bibfield  {author} {\bibinfo {author} {\bibfnamefont {M.~S.}\ \bibnamefont
  {Rudner}}, \bibinfo {author} {\bibfnamefont {N.~H.}\ \bibnamefont {Lindner}},
  \bibinfo {author} {\bibfnamefont {E.}~\bibnamefont {Berg}},\ and\ \bibinfo
  {author} {\bibfnamefont {M.}~\bibnamefont {Levin}},\ }\bibfield  {title}
  {\bibinfo {title} {Anomalous edge states and the bulk-edge correspondence for
  periodically driven two-dimensional systems},\ }\href@noop {} {\bibfield
  {journal} {\bibinfo  {journal} {Physical Review X}\ }\textbf {\bibinfo
  {volume} {3}},\ \bibinfo {pages} {031005} (\bibinfo {year}
  {2013})}\BibitemShut {NoStop}%
\bibitem [{\citenamefont {Else}\ and\ \citenamefont {Nayak}(2016)}]{Else_2016}%
  \BibitemOpen
  \bibfield  {author} {\bibinfo {author} {\bibfnamefont {D.~V.}\ \bibnamefont
  {Else}}\ and\ \bibinfo {author} {\bibfnamefont {C.}~\bibnamefont {Nayak}},\
  }\bibfield  {title} {\bibinfo {title} {Classification of topological phases
  in periodically driven interacting systems},\ }\bibfield  {journal} {\bibinfo
   {journal} {Physical Review B}\ }\textbf {\bibinfo {volume} {93}},\ \href
  {https://doi.org/10.1103/physrevb.93.201103} {10.1103/physrevb.93.201103}
  (\bibinfo {year} {2016})\BibitemShut {NoStop}%
\bibitem [{\citenamefont {Po}\ \emph {et~al.}(2016)\citenamefont {Po},
  \citenamefont {Fidkowski}, \citenamefont {Morimoto}, \citenamefont {Potter},\
  and\ \citenamefont {Vishwanath}}]{Po_2016}%
  \BibitemOpen
  \bibfield  {author} {\bibinfo {author} {\bibfnamefont {H.~C.}\ \bibnamefont
  {Po}}, \bibinfo {author} {\bibfnamefont {L.}~\bibnamefont {Fidkowski}},
  \bibinfo {author} {\bibfnamefont {T.}~\bibnamefont {Morimoto}}, \bibinfo
  {author} {\bibfnamefont {A.~C.}\ \bibnamefont {Potter}},\ and\ \bibinfo
  {author} {\bibfnamefont {A.}~\bibnamefont {Vishwanath}},\ }\bibfield  {title}
  {\bibinfo {title} {Chiral floquet phases of many-body localized bosons},\
  }\bibfield  {journal} {\bibinfo  {journal} {Physical Review X}\ }\textbf
  {\bibinfo {volume} {6}},\ \href {https://doi.org/10.1103/physrevx.6.041070}
  {10.1103/physrevx.6.041070} (\bibinfo {year} {2016})\BibitemShut {NoStop}%
\bibitem [{\citenamefont {Potter}\ and\ \citenamefont
  {Morimoto}(2017)}]{Potter_2017}%
  \BibitemOpen
  \bibfield  {author} {\bibinfo {author} {\bibfnamefont {A.~C.}\ \bibnamefont
  {Potter}}\ and\ \bibinfo {author} {\bibfnamefont {T.}~\bibnamefont
  {Morimoto}},\ }\bibfield  {title} {\bibinfo {title} {Dynamically enriched
  topological orders in driven two-dimensional systems},\ }\bibfield  {journal}
  {\bibinfo  {journal} {Physical Review B}\ }\textbf {\bibinfo {volume} {95}},\
  \href {https://doi.org/10.1103/physrevb.95.155126}
  {10.1103/physrevb.95.155126} (\bibinfo {year} {2017})\BibitemShut {NoStop}%
\bibitem [{\citenamefont {Roy}\ and\ \citenamefont {Harper}(2017)}]{Roy_2017}%
  \BibitemOpen
  \bibfield  {author} {\bibinfo {author} {\bibfnamefont {R.}~\bibnamefont
  {Roy}}\ and\ \bibinfo {author} {\bibfnamefont {F.}~\bibnamefont {Harper}},\
  }\bibfield  {title} {\bibinfo {title} {Floquet topological phases with
  symmetry in all dimensions},\ }\bibfield  {journal} {\bibinfo  {journal}
  {Physical Review B}\ }\textbf {\bibinfo {volume} {95}},\ \href
  {https://doi.org/10.1103/physrevb.95.195128} {10.1103/physrevb.95.195128}
  (\bibinfo {year} {2017})\BibitemShut {NoStop}%
\bibitem [{\citenamefont {Po}\ \emph {et~al.}(2017)\citenamefont {Po},
  \citenamefont {Fidkowski}, \citenamefont {Vishwanath},\ and\ \citenamefont
  {Potter}}]{Po_2017}%
  \BibitemOpen
  \bibfield  {author} {\bibinfo {author} {\bibfnamefont {H.~C.}\ \bibnamefont
  {Po}}, \bibinfo {author} {\bibfnamefont {L.}~\bibnamefont {Fidkowski}},
  \bibinfo {author} {\bibfnamefont {A.}~\bibnamefont {Vishwanath}},\ and\
  \bibinfo {author} {\bibfnamefont {A.~C.}\ \bibnamefont {Potter}},\ }\bibfield
   {title} {\bibinfo {title} {Radical chiral floquet phases in a periodically
  driven kitaev model and beyond},\ }\bibfield  {journal} {\bibinfo  {journal}
  {Physical Review B}\ }\textbf {\bibinfo {volume} {96}},\ \href
  {https://doi.org/10.1103/physrevb.96.245116} {10.1103/physrevb.96.245116}
  (\bibinfo {year} {2017})\BibitemShut {NoStop}%
\bibitem [{\citenamefont {Fidkowski}\ \emph {et~al.}(2019)\citenamefont
  {Fidkowski}, \citenamefont {Po}, \citenamefont {Potter},\ and\ \citenamefont
  {Vishwanath}}]{Fidkowski_2019}%
  \BibitemOpen
  \bibfield  {author} {\bibinfo {author} {\bibfnamefont {L.}~\bibnamefont
  {Fidkowski}}, \bibinfo {author} {\bibfnamefont {H.~C.}\ \bibnamefont {Po}},
  \bibinfo {author} {\bibfnamefont {A.~C.}\ \bibnamefont {Potter}},\ and\
  \bibinfo {author} {\bibfnamefont {A.}~\bibnamefont {Vishwanath}},\ }\bibfield
   {title} {\bibinfo {title} {Interacting invariants for floquet phases of
  fermions in two dimensions},\ }\bibfield  {journal} {\bibinfo  {journal}
  {Physical Review B}\ }\textbf {\bibinfo {volume} {99}},\ \href
  {https://doi.org/10.1103/physrevb.99.085115} {10.1103/physrevb.99.085115}
  (\bibinfo {year} {2019})\BibitemShut {NoStop}%
\bibitem [{\citenamefont {Duschatko}\ \emph {et~al.}(2018)\citenamefont
  {Duschatko}, \citenamefont {Dumitrescu},\ and\ \citenamefont
  {Potter}}]{Duschatko_2018}%
  \BibitemOpen
  \bibfield  {author} {\bibinfo {author} {\bibfnamefont {B.~R.}\ \bibnamefont
  {Duschatko}}, \bibinfo {author} {\bibfnamefont {P.~T.}\ \bibnamefont
  {Dumitrescu}},\ and\ \bibinfo {author} {\bibfnamefont {A.~C.}\ \bibnamefont
  {Potter}},\ }\bibfield  {title} {\bibinfo {title} {Tracking the quantized
  information transfer at the edge of a chiral floquet phase},\ }\bibfield
  {journal} {\bibinfo  {journal} {Physical Review B}\ }\textbf {\bibinfo
  {volume} {98}},\ \href {https://doi.org/10.1103/physrevb.98.054309}
  {10.1103/physrevb.98.054309} (\bibinfo {year} {2018})\BibitemShut {NoStop}%
\bibitem [{\citenamefont {Zhang}\ and\ \citenamefont
  {Levin}(2021)}]{zhang2021classification}%
  \BibitemOpen
  \bibfield  {author} {\bibinfo {author} {\bibfnamefont {C.}~\bibnamefont
  {Zhang}}\ and\ \bibinfo {author} {\bibfnamefont {M.}~\bibnamefont {Levin}},\
  }\bibfield  {title} {\bibinfo {title} {Classification of interacting floquet
  phases with u (1) symmetry in two dimensions},\ }\href@noop {} {\bibfield
  {journal} {\bibinfo  {journal} {Physical Review B}\ }\textbf {\bibinfo
  {volume} {103}},\ \bibinfo {pages} {064302} (\bibinfo {year}
  {2021})}\BibitemShut {NoStop}%
\bibitem [{\citenamefont {Zhang}\ and\ \citenamefont
  {Levin}(2022)}]{zhang2022bulk}%
  \BibitemOpen
  \bibfield  {author} {\bibinfo {author} {\bibfnamefont {C.}~\bibnamefont
  {Zhang}}\ and\ \bibinfo {author} {\bibfnamefont {M.}~\bibnamefont {Levin}},\
  }\bibfield  {title} {\bibinfo {title} {Bulk-boundary correspondence for
  interacting floquet systems in two dimensions},\ }\href@noop {} {\bibfield
  {journal} {\bibinfo  {journal} {arXiv preprint arXiv:2209.03975}\ } (\bibinfo
  {year} {2022})}\BibitemShut {NoStop}%
\bibitem [{\citenamefont {Harper}\ \emph {et~al.}(2020)\citenamefont {Harper},
  \citenamefont {Roy}, \citenamefont {Rudner},\ and\ \citenamefont
  {Sondhi}}]{Harper_2020}%
  \BibitemOpen
  \bibfield  {author} {\bibinfo {author} {\bibfnamefont {F.}~\bibnamefont
  {Harper}}, \bibinfo {author} {\bibfnamefont {R.}~\bibnamefont {Roy}},
  \bibinfo {author} {\bibfnamefont {M.~S.}\ \bibnamefont {Rudner}},\ and\
  \bibinfo {author} {\bibfnamefont {S.}~\bibnamefont {Sondhi}},\ }\bibfield
  {title} {\bibinfo {title} {Topology and broken symmetry in floquet systems},\
  }\href {https://doi.org/10.1146/annurev-conmatphys-031218-013721} {\bibfield
  {journal} {\bibinfo  {journal} {Annual Review of Condensed Matter Physics}\
  }\textbf {\bibinfo {volume} {11}},\ \bibinfo {pages} {345} (\bibinfo {year}
  {2020})}\BibitemShut {NoStop}%
\bibitem [{\citenamefont {De~Roeck}\ and\ \citenamefont
  {Huveneers}(2017)}]{de2017stability}%
  \BibitemOpen
  \bibfield  {author} {\bibinfo {author} {\bibfnamefont {W.}~\bibnamefont
  {De~Roeck}}\ and\ \bibinfo {author} {\bibfnamefont {F.}~\bibnamefont
  {Huveneers}},\ }\bibfield  {title} {\bibinfo {title} {Stability and
  instability towards delocalization in many-body localization systems},\
  }\href@noop {} {\bibfield  {journal} {\bibinfo  {journal} {Physical Review
  B}\ }\textbf {\bibinfo {volume} {95}},\ \bibinfo {pages} {155129} (\bibinfo
  {year} {2017})}\BibitemShut {NoStop}%
\bibitem [{\citenamefont {Dubail}\ and\ \citenamefont
  {Read}(2015)}]{Dubail_2015}%
  \BibitemOpen
  \bibfield  {author} {\bibinfo {author} {\bibfnamefont {J.}~\bibnamefont
  {Dubail}}\ and\ \bibinfo {author} {\bibfnamefont {N.}~\bibnamefont {Read}},\
  }\bibfield  {title} {\bibinfo {title} {Tensor network trial states for chiral
  topological phases in two dimensions and a no-go theorem in any dimension},\
  }\bibfield  {journal} {\bibinfo  {journal} {Physical Review B}\ }\textbf
  {\bibinfo {volume} {92}},\ \href {https://doi.org/10.1103/physrevb.92.205307}
  {10.1103/physrevb.92.205307} (\bibinfo {year} {2015})\BibitemShut {NoStop}%
\bibitem [{\citenamefont {Tantivasadakarn}\ \emph {et~al.}(2021)\citenamefont
  {Tantivasadakarn}, \citenamefont {Thorngren}, \citenamefont {Vishwanath},\
  and\ \citenamefont {Verresen}}]{tantivasadakarn2021long}%
  \BibitemOpen
  \bibfield  {author} {\bibinfo {author} {\bibfnamefont {N.}~\bibnamefont
  {Tantivasadakarn}}, \bibinfo {author} {\bibfnamefont {R.}~\bibnamefont
  {Thorngren}}, \bibinfo {author} {\bibfnamefont {A.}~\bibnamefont
  {Vishwanath}},\ and\ \bibinfo {author} {\bibfnamefont {R.}~\bibnamefont
  {Verresen}},\ }\bibfield  {title} {\bibinfo {title} {Long-range entanglement
  from measuring symmetry-protected topological phases},\ }\href@noop {}
  {\bibfield  {journal} {\bibinfo  {journal} {arXiv preprint arXiv:2112.01519}\
  } (\bibinfo {year} {2021})}\BibitemShut {NoStop}%
\bibitem [{\citenamefont {Verresen}\ \emph {et~al.}(2021)\citenamefont
  {Verresen}, \citenamefont {Tantivasadakarn},\ and\ \citenamefont
  {Vishwanath}}]{verresen2021efficiently}%
  \BibitemOpen
  \bibfield  {author} {\bibinfo {author} {\bibfnamefont {R.}~\bibnamefont
  {Verresen}}, \bibinfo {author} {\bibfnamefont {N.}~\bibnamefont
  {Tantivasadakarn}},\ and\ \bibinfo {author} {\bibfnamefont {A.}~\bibnamefont
  {Vishwanath}},\ }\bibfield  {title} {\bibinfo {title} {Efficiently preparing
  ghz, topological and fracton states by measuring cold atoms},\ }\href@noop {}
  {\bibfield  {journal} {\bibinfo  {journal} {arXiv preprint arXiv:2112.03061}\
  } (\bibinfo {year} {2021})}\BibitemShut {NoStop}%
\bibitem [{\citenamefont {Zhang}\ \emph
  {et~al.}(2022{\natexlab{a}})\citenamefont {Zhang}, \citenamefont {Aasen},\
  and\ \citenamefont {Vijay}}]{zhang2022x}%
  \BibitemOpen
  \bibfield  {author} {\bibinfo {author} {\bibfnamefont {Z.}~\bibnamefont
  {Zhang}}, \bibinfo {author} {\bibfnamefont {D.}~\bibnamefont {Aasen}},\ and\
  \bibinfo {author} {\bibfnamefont {S.}~\bibnamefont {Vijay}},\ }\bibfield
  {title} {\bibinfo {title} {The x-cube floquet code},\ }\href@noop {}
  {\bibfield  {journal} {\bibinfo  {journal} {arXiv preprint arXiv:2211.05784}\
  } (\bibinfo {year} {2022}{\natexlab{a}})}\BibitemShut {NoStop}%
\bibitem [{\citenamefont {Dua}\ \emph {et~al.}(2023)\citenamefont {Dua},
  \citenamefont {Ellison}, \citenamefont {Sullivan},\ and\ \citenamefont
  {Tantivasadakarn}}]{Arpit2023}%
  \BibitemOpen
  \bibfield  {author} {\bibinfo {author} {\bibfnamefont {A.}~\bibnamefont
  {Dua}}, \bibinfo {author} {\bibfnamefont {T.}~\bibnamefont {Ellison}},
  \bibinfo {author} {\bibfnamefont {J.}~\bibnamefont {Sullivan}},\ and\
  \bibinfo {author} {\bibfnamefont {N.}~\bibnamefont {Tantivasadakarn}},\
  }\bibfield  {title} {\bibinfo {title} {Topological floquet codes: new
  examples from general principles},\ }\href@noop {} {\bibfield  {journal}
  {\bibinfo  {journal} {In preparation}\ } (\bibinfo {year}
  {2023})}\BibitemShut {NoStop}%
\bibitem [{\citenamefont {Potter}\ and\ \citenamefont
  {Vasseur}(2016)}]{Potter_2016}%
  \BibitemOpen
  \bibfield  {author} {\bibinfo {author} {\bibfnamefont {A.~C.}\ \bibnamefont
  {Potter}}\ and\ \bibinfo {author} {\bibfnamefont {R.}~\bibnamefont
  {Vasseur}},\ }\bibfield  {title} {\bibinfo {title} {Symmetry constraints on
  many-body localization},\ }\bibfield  {journal} {\bibinfo  {journal}
  {Physical Review B}\ }\textbf {\bibinfo {volume} {94}},\ \href
  {https://doi.org/10.1103/physrevb.94.224206} {10.1103/physrevb.94.224206}
  (\bibinfo {year} {2016})\BibitemShut {NoStop}%
\bibitem [{\citenamefont {Wen}(2003)}]{WenPlaquette}%
  \BibitemOpen
  \bibfield  {author} {\bibinfo {author} {\bibfnamefont {X.-G.}\ \bibnamefont
  {Wen}},\ }\bibfield  {title} {\bibinfo {title} {Quantum orders in an exact
  soluble model},\ }\href {https://doi.org/10.1103/PhysRevLett.90.016803}
  {\bibfield  {journal} {\bibinfo  {journal} {Phys. Rev. Lett.}\ }\textbf
  {\bibinfo {volume} {90}},\ \bibinfo {pages} {016803} (\bibinfo {year}
  {2003})}\BibitemShut {NoStop}%
\bibitem [{\citenamefont {Bombin}\ \emph {et~al.}(2009)\citenamefont {Bombin},
  \citenamefont {Kargarian},\ and\ \citenamefont
  {Martin-Delgado}}]{Bombin2009fermions}%
  \BibitemOpen
  \bibfield  {author} {\bibinfo {author} {\bibfnamefont {H.}~\bibnamefont
  {Bombin}}, \bibinfo {author} {\bibfnamefont {M.}~\bibnamefont {Kargarian}},\
  and\ \bibinfo {author} {\bibfnamefont {M.~A.}\ \bibnamefont
  {Martin-Delgado}},\ }\bibfield  {title} {\bibinfo {title} {Interacting
  anyonic fermions in a two-body color code model},\ }\bibfield  {journal}
  {\bibinfo  {journal} {Physical Review B}\ }\textbf {\bibinfo {volume} {80}},\
  \href {https://doi.org/10.1103/physrevb.80.075111}
  {10.1103/physrevb.80.075111} (\bibinfo {year} {2009})\BibitemShut {NoStop}%
\bibitem [{\citenamefont {Barkeshli}\ \emph {et~al.}(2013)\citenamefont
  {Barkeshli}, \citenamefont {Jian},\ and\ \citenamefont
  {Qi}}]{Barkeshli_2013}%
  \BibitemOpen
  \bibfield  {author} {\bibinfo {author} {\bibfnamefont {M.}~\bibnamefont
  {Barkeshli}}, \bibinfo {author} {\bibfnamefont {C.-M.}\ \bibnamefont
  {Jian}},\ and\ \bibinfo {author} {\bibfnamefont {X.-L.}\ \bibnamefont {Qi}},\
  }\bibfield  {title} {\bibinfo {title} {Twist defects and projective
  non-abelian braiding statistics},\ }\bibfield  {journal} {\bibinfo  {journal}
  {Physical Review B}\ }\textbf {\bibinfo {volume} {87}},\ \href
  {https://doi.org/10.1103/physrevb.87.045130} {10.1103/physrevb.87.045130}
  (\bibinfo {year} {2013})\BibitemShut {NoStop}%
\bibitem [{\citenamefont {Barkeshli}\ \emph {et~al.}(2019)\citenamefont
  {Barkeshli}, \citenamefont {Bonderson}, \citenamefont {Cheng},\ and\
  \citenamefont {Wang}}]{Cheng_defects_2019}%
  \BibitemOpen
  \bibfield  {author} {\bibinfo {author} {\bibfnamefont {M.}~\bibnamefont
  {Barkeshli}}, \bibinfo {author} {\bibfnamefont {P.}~\bibnamefont
  {Bonderson}}, \bibinfo {author} {\bibfnamefont {M.}~\bibnamefont {Cheng}},\
  and\ \bibinfo {author} {\bibfnamefont {Z.}~\bibnamefont {Wang}},\ }\bibfield
  {title} {\bibinfo {title} {Symmetry fractionalization, defects, and gauging
  of topological phases},\ }\href {https://doi.org/10.1103/PhysRevB.100.115147}
  {\bibfield  {journal} {\bibinfo  {journal} {Phys. Rev. B}\ }\textbf {\bibinfo
  {volume} {100}},\ \bibinfo {pages} {115147} (\bibinfo {year}
  {2019})}\BibitemShut {NoStop}%
\bibitem [{\citenamefont {Else}\ and\ \citenamefont
  {Thorngren}(2019)}]{else2019crystalline}%
  \BibitemOpen
  \bibfield  {author} {\bibinfo {author} {\bibfnamefont {D.~V.}\ \bibnamefont
  {Else}}\ and\ \bibinfo {author} {\bibfnamefont {R.}~\bibnamefont
  {Thorngren}},\ }\bibfield  {title} {\bibinfo {title} {Crystalline topological
  phases as defect networks},\ }\href@noop {} {\bibfield  {journal} {\bibinfo
  {journal} {Physical Review B}\ }\textbf {\bibinfo {volume} {99}},\ \bibinfo
  {pages} {115116} (\bibinfo {year} {2019})}\BibitemShut {NoStop}%
\bibitem [{\citenamefont {Barkeshli}\ \emph {et~al.}(2015)\citenamefont
  {Barkeshli}, \citenamefont {Jiang}, \citenamefont {Thomale},\ and\
  \citenamefont {Qi}}]{Barkeshli_2015}%
  \BibitemOpen
  \bibfield  {author} {\bibinfo {author} {\bibfnamefont {M.}~\bibnamefont
  {Barkeshli}}, \bibinfo {author} {\bibfnamefont {H.-C.}\ \bibnamefont
  {Jiang}}, \bibinfo {author} {\bibfnamefont {R.}~\bibnamefont {Thomale}},\
  and\ \bibinfo {author} {\bibfnamefont {X.-L.}\ \bibnamefont {Qi}},\
  }\bibfield  {title} {\bibinfo {title} {Generalized kitaev models and
  extrinsic non-abelian twist defects},\ }\bibfield  {journal} {\bibinfo
  {journal} {Physical Review Letters}\ }\textbf {\bibinfo {volume} {114}},\
  \href {https://doi.org/10.1103/physrevlett.114.026401}
  {10.1103/physrevlett.114.026401} (\bibinfo {year} {2015})\BibitemShut
  {NoStop}%
\bibitem [{\citenamefont {Zhang}\ \emph
  {et~al.}(2022{\natexlab{b}})\citenamefont {Zhang}, \citenamefont {Aasen},\
  and\ \citenamefont {Vijay}}]{XcubeFloquet_2022}%
  \BibitemOpen
  \bibfield  {author} {\bibinfo {author} {\bibfnamefont {Z.}~\bibnamefont
  {Zhang}}, \bibinfo {author} {\bibfnamefont {D.}~\bibnamefont {Aasen}},\ and\
  \bibinfo {author} {\bibfnamefont {S.}~\bibnamefont {Vijay}},\ }\bibfield
  {title} {\bibinfo {title} {The x-cube floquet code}\ }\href
  {https://doi.org/10.48550/ARXIV.2211.05784} {10.48550/ARXIV.2211.05784}
  (\bibinfo {year} {2022}{\natexlab{b}})\BibitemShut {NoStop}%
\bibitem [{\citenamefont {Williamson}(2022)}]{Williamson2022spacetime}%
  \BibitemOpen
  \bibfield  {author} {\bibinfo {author} {\bibfnamefont {D.~J.}\ \bibnamefont
  {Williamson}},\ }\bibfield  {title} {\bibinfo {title} {Spacetime topological
  defect networks and floquet codes}} (\bibinfo {year} {2022}),\ \bibinfo
  {note} {{KITP} Conference: Noisy Intermediate-Scale Quantum Systems: Advances
  and Applications}\BibitemShut {NoStop}%
\bibitem [{\citenamefont {Vu}\ \emph {et~al.}(2023)\citenamefont {Vu},
  \citenamefont {Lavasani}, \citenamefont {Lee},\ and\ \citenamefont
  {Fisher}}]{Fisher2023}%
  \BibitemOpen
  \bibfield  {author} {\bibinfo {author} {\bibfnamefont {D.}~\bibnamefont
  {Vu}}, \bibinfo {author} {\bibfnamefont {A.}~\bibnamefont {Lavasani}},
  \bibinfo {author} {\bibfnamefont {J.~Y.}\ \bibnamefont {Lee}},\ and\ \bibinfo
  {author} {\bibfnamefont {M.~P.~A.}\ \bibnamefont {Fisher}},\ }\href@noop {}
  {\bibinfo {title} {Measurement-induced floquet enriched topological order}}
  (\bibinfo {year} {2023}),\ \Eprint {https://arxiv.org/abs/2303.01533}
  {arXiv:2303.01533 [quant-ph]} \BibitemShut {NoStop}%
\bibitem [{\citenamefont {Gidney}\ \emph {et~al.}(2021)\citenamefont {Gidney},
  \citenamefont {Newman}, \citenamefont {Fowler},\ and\ \citenamefont
  {Broughton}}]{Gidney2021faulttolerant}%
  \BibitemOpen
  \bibfield  {author} {\bibinfo {author} {\bibfnamefont {C.}~\bibnamefont
  {Gidney}}, \bibinfo {author} {\bibfnamefont {M.}~\bibnamefont {Newman}},
  \bibinfo {author} {\bibfnamefont {A.}~\bibnamefont {Fowler}},\ and\ \bibinfo
  {author} {\bibfnamefont {M.}~\bibnamefont {Broughton}},\ }\bibfield  {title}
  {\bibinfo {title} {A {F}ault-{T}olerant {H}oneycomb {M}emory},\ }\href
  {https://doi.org/10.22331/q-2021-12-20-605} {\bibfield  {journal} {\bibinfo
  {journal} {{Quantum}}\ }\textbf {\bibinfo {volume} {5}},\ \bibinfo {pages}
  {605} (\bibinfo {year} {2021})}\BibitemShut {NoStop}%
\bibitem [{\citenamefont {Ellison}\ \emph {et~al.}(2023)\citenamefont
  {Ellison}, \citenamefont {Sullivan}, \citenamefont {Dua},\ and\ \citenamefont
  {Tantivasadakarn}}]{FCtwist_2023}%
  \BibitemOpen
  \bibfield  {author} {\bibinfo {author} {\bibfnamefont {T.}~\bibnamefont
  {Ellison}}, \bibinfo {author} {\bibfnamefont {J.}~\bibnamefont {Sullivan}},
  \bibinfo {author} {\bibfnamefont {A.}~\bibnamefont {Dua}},\ and\ \bibinfo
  {author} {\bibfnamefont {N.}~\bibnamefont {Tantivasadakarn}},\ }\bibfield
  {title} {\bibinfo {title} {Floquet codes with a twist},\ }\href@noop {}
  {\bibfield  {journal} {\bibinfo  {journal} {In preparation}\ } (\bibinfo
  {year} {2023})}\BibitemShut {NoStop}%
\bibitem [{\citenamefont {Roberts}\ \emph {et~al.}(2023)\citenamefont
  {Roberts}, \citenamefont {Vijay}, \citenamefont {Vishwanath},\ and\
  \citenamefont {Dua}}]{roberts2023geometric}%
  \BibitemOpen
  \bibfield  {author} {\bibinfo {author} {\bibfnamefont {B.}~\bibnamefont
  {Roberts}}, \bibinfo {author} {\bibfnamefont {S.}~\bibnamefont {Vijay}},
  \bibinfo {author} {\bibfnamefont {A.}~\bibnamefont {Vishwanath}},\ and\
  \bibinfo {author} {\bibfnamefont {A.}~\bibnamefont {Dua}},\ }\bibfield
  {title} {\bibinfo {title} {Topological invariants of floquet codes},\
  }\href@noop {} {\bibfield  {journal} {\bibinfo  {journal} {In preparation}\ }
  (\bibinfo {year} {2023})}\BibitemShut {NoStop}%
\bibitem [{\citenamefont {Aasen}\ \emph {et~al.}(2023)\citenamefont {Aasen},
  \citenamefont {Haah}, \citenamefont {Li},\ and\ \citenamefont
  {Mong}}]{Dave2023}%
  \BibitemOpen
  \bibfield  {author} {\bibinfo {author} {\bibfnamefont {D.}~\bibnamefont
  {Aasen}}, \bibinfo {author} {\bibfnamefont {J.}~\bibnamefont {Haah}},
  \bibinfo {author} {\bibfnamefont {Z.}~\bibnamefont {Li}},\ and\ \bibinfo
  {author} {\bibfnamefont {R.~S.}\ \bibnamefont {Mong}},\ }\bibfield  {title}
  {\bibinfo {title} {Anomalous boundary actions of measurement dynamics},\
  }\href@noop {} {\bibfield  {journal} {\bibinfo  {journal} {In preparation}\ }
  (\bibinfo {year} {2023})}\BibitemShut {NoStop}%
\bibitem [{\citenamefont {Kitaev}(2001)}]{Kitaev_2001}%
  \BibitemOpen
  \bibfield  {author} {\bibinfo {author} {\bibfnamefont {A.~Y.}\ \bibnamefont
  {Kitaev}},\ }\bibfield  {title} {\bibinfo {title} {Unpaired majorana fermions
  in quantum wires},\ }\href {https://doi.org/10.1070/1063-7869/44/10s/s29}
  {\bibfield  {journal} {\bibinfo  {journal} {Physics-Uspekhi}\ }\textbf
  {\bibinfo {volume} {44}},\ \bibinfo {pages} {131} (\bibinfo {year}
  {2001})}\BibitemShut {NoStop}%
\end{thebibliography}%


%

\appendix

\section{Chiral Unitary Index\label{chiral index}}
In this Appendix, we briefly summarize the chiral unitary index for fermionic systems, such as the gauged fermion system relevant to the unitary-lift of the honeycomb Floquet code (HFC) model. For details, we refer the reader to~\cite{Fidkowski_2019}. In this appendix, we consider freezing the $\Z_2$ gauge fields and working with a fermion system in the fixed gauge background. 
Fermionic systems are formally described by a $\Z_2$-graded tensor product Hilbert space, where the $\Z_2$-grading simply associates each state, $|i\>$ with a $\Z_2$ number $|i|=\pm 1$ that indicates whether there are an even or odd fermion parity. Furthermore, the graded tensor product of different fermionic subsystems is then defined as\footnote{Formally speaking, there is a natural isomorphism on $\Z_2$-graded tensor product space: $\F:~V\otimes_g W\to W\otimes_g V, |i\>\otimes_g|j\> \mapsto (-1)^{|i||j|}|j\>\otimes_g |i\>$}
$|i\>\otimes_g |j\> = (-1)^{|i|\cdot|j|}|j\>\otimes_g |i\>$.

The index applies to $2d$ dynamics generated by a local Hamiltonian evolution $U(t) = \mathcal{T}e^{-iHt}$ induces a locality-preserving unitary evolution, such that the time evolution for one Floquet period can be factorized into bulk and edge components: 
\begin{align}
U=U(t=1) = U_\text{bulk} \otimes_g U_\text{edge}.
\end{align} 

Crucially, while $U$ is generated by a local $2d$ Hamiltonian, in topologically non-trivial cases, $U_\text{edge}$ will not be generated by any edge-local Hamiltonian. That is, while $U_\text{edge}$ is locality preserving, i.e. it maps (quasi)local operators to other nearby (quasi)-local operators, it need not be locally-generated.
The chiral index $\nu(U) = \nu(U_\text{edge})$ measures the amount of quantum information transported along the edge by $U_\text{edge}$.
This index obeys the multiplicative composition rules: 
\begin{align}
\nu(U\otimes_g V) = \nu(U)\nu(V) = \nu(UV).
\end{align}
$\nu(U)=1$ signifies a topologically-trivial unitary, and is satisfied iff $U_\text{edge}$ is locally generated, i.e. iff $U_\text{edge} = \mathcal{T}e^{-i\int_0^1 H_{1d}(t)dt}=1$ for some $1d$-local $H_{1d(t)}$.

While the total fermion parity of $U$ is necessarily even, $U_\text{bulk/edge}$ could individually have even or odd fermion parity. If $U_\text{edge}$ always has even-fermion parity, then $\nu(U_\text{edge})\in \mathbb{Q}$ takes rational values. On the other hand, $\nu(U_\text{edge})$ takes an irrational value, $\sqrt{2}\mathbb{Q}$, iff $U_\text{edge}$ has opposite fermion parity for periodic vs anti-periodic boundary conditions (equivalently if $U_\text{edge}$ changes fermion parity in response to inserting a $\Z_2$ fermion parity flux)~\cite{Kitaev_2001}.

Absent additional symmetries, $\nu$ represents a complete classification of $2d$ unitary loops and MBL dynamics of non-fractionalized fermion systems or $\Z_2$ topological orders with emergent fermions~\cite{Po_2017,Fidkowski_2019}.

In this section we focus only on $1d$ locality preserving unitaries $U$, such as $U_\text{edge}$ defined above, and for convenience we drop the ``edge" subscript in subsequent expressions.

The Chiral unitary index was first formulated by GNVW~\cite{Gross_2012} for bosonic systems and later generalized in~\cite{Fidkowski_2019,Po_2017} to fermion systems. The original formulation of this index is in terms of overlaps between operator algebras $\mathcal{A}$, $\mathcal{B}$ of observables. In the fermion context, we can think of the operator algebras $\mathcal{A}$ for a region $A$ as the algebra generated polynomials of the $2|A|$ Majorana modes in region $A$ and products thereof, where $|A|$ denotes the number of sites in region $A$. Then the overlap between two such algebras is defined as:
\begin{align}
\<\mathcal{A},\mathcal{B}\> = 2^{-|A\cup B|} \sqrt{
\sum_{a=1}^{2^{2|A|}}\sum_{b=1}^{2^{2|B|}} \[\text{tr}_{A\cup B} e_a^\dagger e_b\]|^2
}
\end{align}
where $e_{a,b}$ denote an othonormal basis of operators for $\mathcal{A,B}$ respectively. This definition has the property that overlap of $\mathcal{A}$ with itself satisfies $\<\mathcal{A},\mathcal{A}\> = 2^{|A|}$, whereas the overlap of commuting (in the $\Z_2$-graded sense) algebras $\mathcal{A,B}$ is $\<\mathcal{A,B}\> = 1$.

The chiral unitary (a.k.a. GNVW) index is then defined as follows: Take regions $A$ and $B$ to be abutting (at, say $x=0$) but non-overlapping intervals that are sufficiently large such that operators near the interface cannot spread or move outside of $A\cup B$ during one period. Then, in terms of the associated observable algebras $\mathcal{A,B}$ the chiral unitary index:
\begin{align}
\nu(U) = \frac{\<U^\dagger \mathcal{A} U,\mathcal{B}\>}{\<U^\dagger \mathcal{B} U,\mathcal{A}\>}
\end{align}
keeps track of of the ratio of how many operators flow from $A\rightarrow B$ versus the amount that flow from $B\rightarrow A$ under the evolution of $U$ (or equivalently in the Schrodinger picture of the evolution, how many states of in region $B$ flow into region $A$ under the evolution). 

For the Majorana translation dynamics realized at the edge of the unitary circuit version of the HFC code above (see Fig.~\ref{fig:Unitary action}), it suffices to choose regions $A$ and $B$ of the edge that only contain two Majorana modes each, such that $\mathcal{A} = \{c_1,c_2,c_1c_2\}$ and $\mathcal{B} = \{c_3,c_4,c_3c_4\}$. Then, under the Majorana translation the only nontrivial overlap between from the contribution to $\<\mathcal{A,B}\>$ from $U^\dagger c_2 U = c_3$ (all other operators evolve purely within $A$ or $B$, or move from $B$ to outside of $A\cup B$). This gives $\nu(U) = \sqrt{2}$.

The generalized HFC with general twist defects correspond to similar unitary models in which the bulk defects are swapped around short loops and the boundary defects get translated along the edge, with the translation of the orientation depending on the convention for turning the generalized HFC into a unitary loop. While there is not a rigorous theory of QCAs for general anyonic degrees of freedom, a natural generalization of the chiral Floquet index would then give $\nu(U) = d_\sigma \mathbb{Q}$, where $d_\sigma$ is the quantum dimension of the twist defect.

\section{Measurements in continuum defect network\label{app: GHFC measurement scheme}}

\begin{figure}[h]
    \centering
    \includegraphics[width = 1\columnwidth]{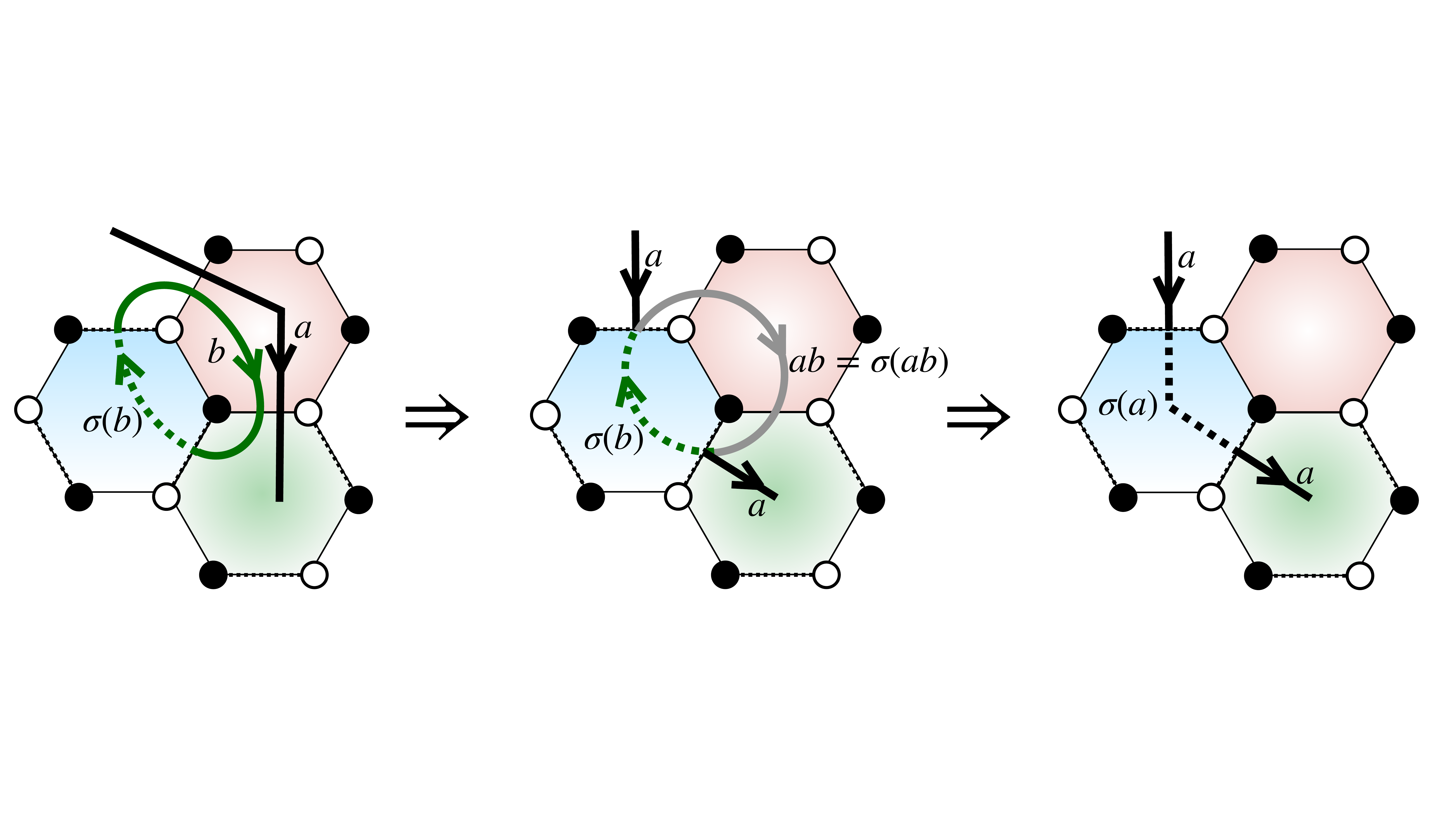}
\caption{{\bf ``Moving" logical operators across twist defects -- } $ab$ is a invariant anyon. One can use the measured $b$ check to move an $a$ string by first fusing the $a$ string with the $b$-string outside the blue plaquette to obtain an invariant anyon $ab$, and then deforming the $ab$ string into the blue plaquette and fuse it with the $\sigma(\overline{b})$-string to obtain a temporary $\sigma(a)$-string.}
    \label{fig: transparent anyon}
\end{figure}
Here we provide details on how to determine minimal set of measurements required to produce general Floquet codes discussed in Sec. \ref{sec: General twist defects}. Note that measuring braiding checks for all anyons will definitely be sufficient for generating the logical movements, in many cases it is not necessary and reduction of number of measurements is possible. For instance, 1). When the braiding checks of $a_i$ are known, braiding checks of any anyon that can be generated by $a_i$s is also known. 2). When an anyon $ab$ is invariant under $\sigma$, measuring the braiding of one of them suffice to generate logical transform for both $a$-strings and $b$-strings. The process is shown in Fig.~\ref{fig: transparent anyon}. Before we discuss the general scheme for measurement reduction based on the above two rules, let us consider 2 helpful examples. We will use $\mathcal{D}_{\mathbb{Z}_N}$, the quantum double of $\mathbb{Z}_N$, to denote the $\mathbb{Z}_N$ toric code.

\subsubsection{$\mathcal{D}_{\mathbb{Z}_N}$ with $e \to m^q, m \to e^p$}
\label{sec: ZN em}
In this case one only needs to measure braiding check of $e$ in every round. To see this, notice firstly the anyon $em^q$ is invariant, therefore one can move an $m^q$-string using $e$-checks by rule 2). Since an $m$-string is the $p$-th power of an $m^q$-string: $(m^q)^p=1$, according to rule 1) $m$-string is now also movable. Since $e,m$ generates the toric code, all anyon strings are now movable.

\subsubsection{$(\mathcal{D}_{\mathbb{Z}_N})^3$ with $(e_i,m_i) \to (e_{i+1},m_{i+1})$}
\label{sec:ZN ee}
In this case one only needs to measure braiding of $e_1,e_2,m_1,m_2$. The invariant anyons are $e_1e_2e_3$ and $m_1m_2m_3$, therefore It is clear $e_3$-strings can be moved using $e_1,e_2$ checks, and $m_3$-checks can be moved using $m_1,m_2$-checks. No further reduction of measurements can be performed, since now no two anyons in the set $\{e_1,e_2,m_1,m_2\}$ are related by any of the two rules.


%

\subsubsection{General scheme for measurement reduction}
With the just discussed examples in hand we now give a recipe for what braiding checks to measure in the general case.

Let $\{a_1,\cdots,a_N\}$ be the collection of generators of the Abelian topological order of interests and start by considering the trajectory of $a_1$ under the permutation $\sigma$: $a_1,\sigma(a_1),\cdots, \sigma^{k_1-1}(a_1)$, where $k_1$ is the smallest number such that $\sigma^{k_1}(a_1) = a_1$, i.e. the order of $a_1$ under $\sigma$.  The fusion of all anyons on the trajectory, $a_1\sigma(a_1)\sigma^2(a_1)\cdots \sigma^{k_1-1}(a_1)$, is invariant under $\sigma$, therefore the braiding check of one of them can be thrown out according to rule 2). Let us choose to measure the first $k_1-1$ anyons. Now according to rule 1). any anyon string that is in the group generated by the anyons on the trajectory of $a_1$, denoted as $\mathcal{A}(a_1)$, is now movable. In the next step one searches for any generator $a_{i_2}$ that is still immovable, i.e. $a_{i_2}$ that is not in $\mathcal{A}(a_1)$. One performs the measurements for all anyons on the trajectory of $a_{i_2}$ except $\sigma^{k_2-1}(a_{i_2})$, as we did for $a_{1}$. Then any anyon string that is in the group generated by the trajectories of $a_{i_1},a_{i_2}$ is now known. One can then look for any generator whose braiding is still unknown and repeat this process until we cover all the generators. At the termination of our search we will have used $Q<N$ generators $a_j$ and performed $\sum_{j=1}^{Q} (k_j-1)$ braiding check measurements.

\section{Analysis of the generalized HFC code space\label{app: generalized HFC gapped proof}}
In this section we provide proof that all local degrees of freedom are frozen by the measurements and certain instantaneous code space will emerge after any measurement round $\ge3$ in the parton realization of the generalized HFC models. 

First, the persistent stabilizers $\mP_{[a]}$ on a plaquette mutually commute and form a representation of $\TO_0^\sigma$. The joint eigenvalues of $\mP_{[a]}$: $\{p_{[a]}\}$ can be viewed as a map from $\TO_0^\sigma$ to $U(1)$(since $\mP_{[a]}$s are unitary, their eigenvalues are phase factors), and $\mP_{[a]}$ being an representation means $p_{[a]}$ is an element of the Pontryagin dual of $\TO_0^\sigma$: $\widehat{\TO_0^\sigma}$. It is possible for a joint eigenspace of $\mP_{[a]}$s to be degenerate. In our case, $\mP_{[a]}$s are plaquette operators therefore their joint eigenspace must be degenerate. We prove that the degeneracy associated with different joint eigenvalues $\{p_{[a]}\},\{p'_{[a]}\}$ must be the same.  On a plaquette the operator $\mP_{[a]}$ takes the form 
\begin{align}
    T^a_{1,z}T^{\sigma(a)}_{2,y}T^a_{3,x}T^{\sigma(a)}_{4,z}T^a_{5,y}T^{\sigma(a)}_{6,x}
\end{align}
For any on-site operator $T^a_{i,\alpha}$, its joint eigenvalues, $\psi([a])$, as a function of $[a]$, may be thought of as a map from $\TO_0^\sigma$ to $U(1)$. Moreover, since $T^a_{i,\alpha}$ form a representation of the group $\TO_0^\sigma$, its eigenvalues $\psi([a])$ is an element of $\text{Hom}(\TO_0^\sigma,U(1))=\widehat{\TO_0^\sigma}$, the Pontryagain dual of $\TO_0^\sigma$. In fact, the joint eigenvalues of $T^a_{i,\alpha}$ can be any element of $\widehat{\TO_0^\sigma}$.

\begin{lemma}
Let $\psi\in \widehat{\TO_0^\sigma}$, there is an eigenstate of $T^a_x$ with eigenvalues $\psi(a)$. The same is true for $T^a_z,T^a_y$.
\end{lemma}
\textbf{
Proof:} Let $|\psi\rangle=\sum_{g\in \widehat{\TO_0^\sigma}}c(g)|g\rangle$ be an eigenstate of $T_x^a$ with eigenvalue $\psi$: $T^a_x|\psi\rangle=\psi([a])|\psi\rangle$, we then have :
\begin{align}
   T^a_x|\psi\rangle=\sum_{g\in\widehat{\TO_0^\sigma}}c([g])|[ag]\rangle=\sum_{g\in\widehat{\TO_0^\sigma}} c([a^{-1}g])|[g]\rangle
\end{align}
which gives $c([a^{-1}g])=\psi([a])c([g])$. Define an unitary $V_{\phi}$ for any $\phi\in \widehat{\TO_0^\sigma}$ as $V_{\phi}|[g]\rangle:=\phi([g])|[g]\rangle$, then It is clear that $V_{\overline{\phi}}|\psi\rangle$ will have joint eigenvalue $\phi \psi$ of $T^a_x$:
\begin{align}
    T^a_xV_{\overline{\phi}}|\psi\rangle&=\sum_{[g]\in \widehat{\TO_0^\sigma}} c([g])\overline{\phi}([g])|[ag]\rangle\\
    &=\sum_{[g]\in \widehat{\TO_0^\sigma}} c([a^{-1}g])\overline{\phi}([a^{-1}g])|[g]\rangle\\
    &=\sum_{[g]\in \widehat{\TO_0^\sigma}} \psi([a])c([g])\overline{\phi}([a^{-1}])\overline{\phi}([g])|[g]\rangle\\
    &=\psi(a)\phi([a]) \sum_{[g]\in \widehat{\TO_0^\sigma}} c([g])\overline{\phi}([g])|[g]\rangle\\
    &=\psi\phi(a)V_{\overline{\phi}}|\psi\rangle.
\end{align}
Therefore by acting with $V_\phi$ we can generate eigenstates of $T^a_{x}$ with any eigenvalues in $\widehat{\TO_0^\sigma}$.
For $T^a_z$, its action on the basis is given by:
\begin{align}
    T^a_z|g\rangle=e^{i\theta_{a,g\sigma(\overline{g})}}|g\rangle,
\end{align}
therefore it is diagonal in the basis with joint eigenvalues $\psi_g([a])=e^{i\theta_{a,g\sigma(\overline{g})}}$. $\psi_g\neq \psi_{g'}$ for any $[g]\neq [g']$, since if $\psi_g$ were equal to $\psi_{g'}$, then we would have $\theta_{a,g\sigma(\overline{g})}=\theta_{a,g'\sigma(\overline{g'})}$ for any anyon $a$, leading to $g=g'$ up to invariant anyons, i.e. $[g]=[g']$. Therefore $\psi_g$ s enumerate all $| \widehat{\TO_0^\sigma}|=|\TO_0^\sigma|=D$ eigenvalues of $T^a_z$. $\blacksquare$

The 6 local operators making up $\mP_{[a]}$ commute with each other, say they are diagonalized simultaneously with eigenvalues $\psi_i$, then $\mP_{[a]}$ has joint eigenvalues $\prod_{i=1}^6\psi_i$. Clearly the eigenvalues of $\mP_{[a]}$ enumerate all elements of $\widehat{\TO_0^\sigma}$. Restricting to certain joint eigenspace of $\mP_{[a]}$ with eigenvalues $p_{[a]}$ is equivalent to imposing $\prod_{i=1}^6\psi_i=p$, which can be solved by expressing one of $\psi_i$ in terms of the other 5. Therefore It is clear any joint eigenspace of the plaquette stabilizers $\mP_{[a]}$ has dimension $D^5$. 

At a given round the checks on a bond $T^a_iT^a_j$ also form representation of $\TO_0^\sigma$, similar argument shows restricting to their joint eigen space will reduce the Hilbert space dimension by $1/D$. Now we can count how many DOF are left by the persistent stabilizers and checks: If we have $p$ plaquettes there will be $1/3\times 3p=p$ checks of a given type, therefore the total number of local constraints is $p+p=2p$, but we have exactly $2p$ sites with each site having dimension $D$, thus local dof are frozen by the checks and persistent stabilizers. Notice we have global constraints on the local constraints: product of all plaquette stabilizers is 1, and product of type-i plaquettes and type-i checks is 1. Therefore a nontrivial finite dimensional ICS will be generated by the measurements, which corresponds to some topological order, $\TO$.
\section{Generalized Honeycomb code: examples\label{app: generalized HFC examples}}
In the appendix we provide detailed study of several examples of generalized lattice HFCs following the general construction in section \ref{sec: General twist defects} We will make frequent use of the generalized $\Z_N$ Pauli operators, that act on $N$-level qudits with on-site Hilbert space:
$\mathbb{C}_N=\text{span}\{|j\>: j=0,1,\cdots, N-1\}$:
\begin{align}
   Z|j\>&=e^{i\frac{2\pi j}{N}}|j\>, \nonumber\\
    X|j\>&=|j+1~\mod N\>, \nonumber\\
    Y &= X^\dagger Z^\dagger = X^{N-1}Z^{N-1}.
\end{align}
\subsection{$\TO_0=U(1)_N\times U(1)_N$, $\sigma: a_1\leftrightarrow a_2$}
$\TO_0$ is two layers of $1/N$ Laughlin states. Denote the 1/N Laughlin quasiparticles in the two layers as $\epsilon_1,\epsilon_2$. Anyon $\epsilon_1\epsilon_2$ is invariant, the invariant subgroup is generated by it and is $\mathbb{Z}_N$. The quotient group is $\TO_0/\text{Inv}(\sigma)=\Z_N^2/\Z_N=\Z_N$. $\mH$ is the group algebra over $\mathbb{Z}_N, \mathbb{C}[\mathbb{Z}_N]=\mathbb{C}_N=\text{span}\{|\epsilon_1\rangle,|\epsilon_1^2\rangle,\cdots,|\epsilon_1^N=1\rangle\}$. 

The anyon braiding operators, given by the general recipe \eqref{eq:rep of local ops}, are
\begin{align}
T^{\epsilon_1}_z|\epsilon_1^k\rangle&=e^{i\theta_{\epsilon_1,\epsilon_1^k\overline{\epsilon}^k_2}}|\epsilon_1^k\rangle=e^{\frac{2\pi i k}{N}}|\epsilon_1^k\rangle,\\
 T^{\epsilon_1}_x|\epsilon_1^k\>&=|\epsilon_1^{k+1}\>,
\end{align}
from which we read off $T^{\epsilon_1}_z=Z$ and $T^{\epsilon_1}_x=X$, $T^{\epsilon_1}_y=X^\dagger Z^\dagger=Y$.  This is just the genon theory described in~\cite{Barkeshli_2015}.
\begin{table}[h]
\centering
\setlength{\tabcolsep}{12pt} 
\renewcommand{\arraystretch}{1.1}
\begin{tabular}{|M{2.0cm}|M{2.0cm}|M{2.0cm}|}
\hline
$x$-bond checks & $y$-bond checks & $z$-bond checks \\
\hline
 $ X_iX_j$& $Y_iY_j$ &$Z_iZ_j$\\
 \hline
 \end{tabular}
\label{table: pauli schedule 1}
\end{table}

The $\TO$ generated by the measurement sequence has anyons generated by: $\F_1(\epsilon_1),\F_2(\epsilon_1)$, which are self bosons and mutual statistics: $\theta_{\F_1(\epsilon_1),\F_2(\epsilon_1)}=\theta_{\epsilon_1,\epsilon_1\overline{\epsilon}_2}=\theta_{\epsilon_1,\epsilon_1}=\frac{2\pi}{N}$, therefore we identify them as the $e,m$ particles of a $\mathcal{D}_{\mathbb{Z}_N}$. The measurement induced automorphism  is: $\varphi: \F_1(e_1)\leftrightarrow \F_2(e_1)$, which is then identified as $e\leftrightarrow m$--we have reproduced the HFC code of Haah and Hastings.

\subsection{$\TO_0 =\mathcal{D}_{\mathbb{Z}_N},\sigma: a\to \bar{a}, N=2n>2$}
In this case $e^n,m^n$ are invariant, the invariant subgroup is generated by them and is $\mathbb{Z}_2\times \mathbb{Z}_2$. The quotient group is $\mathcal{D}_{\mathbb{Z}_N}/\text{Inv}(\sigma)=(\mathbb{Z}_N\times \mathbb{Z}_N)/(\mathbb{Z}_2\times \mathbb{Z}_2)=\mathbb{Z}_n\times \mathbb{Z}_n$. $\mH$ is then the group algebra over $\mathbb{Z}_n\times \mathbb{Z}_n$, $\mathbb{C}[\mathbb{Z}_n\times \mathbb{Z}_n]=\mathbb{C}_n\otimes \mathbb{C}_n$ which has dimension $D=n^2$. 

A basis for $\mH$ is $\{|[e^k],[m^l]\rangle|,k,l=0,\cdots, n-1\}$, where $[~]$ is the equivalent class modulo $\text{Inv}(\sigma)$, which satisfies $[e^n]=[m^n]=[1]$. The anyon braiding operators are given by\eqref{eq:rep of local ops} as:
\begin{align}
    T^{e}_z|[e^k],[m^l]\rangle&=e^{i\theta_{e,e^km^l\sigma(\overline{e}^k\overline{m}^l)}}|[e^k],[m^l]\rangle\\
    &=e^{\frac{2\pi i 2l}{N}}|[e^k],[m^l]\rangle=e^{\frac{2\pi i l}{n}}|[e^k],[m^l]\rangle,
\end{align}
from which we read off $T^e_z=Z_2$. Similarly following the recipe \eqref{eq:rep of local ops} we see $T^e_x=X_1$, $T^m_z=Z_1$, $T^m_x=X_2$. Then $T^e_y=T^{e\dagger}_xT^{e\dagger}_z=X_1^\dagger Z_2^\dagger$ and $T^m_y=T^{m\dagger}_xT^{m\dagger}_z=X_2^\dagger Z_1^\dagger$.
\begin{table}[h]
\centering
\setlength{\tabcolsep}{12pt} 
\renewcommand{\arraystretch}{2}
\begin{tabular}{|M{2.0cm}|M{2.0cm}|M{2.0cm}|}
\hline
$x$-bond checks & $y$-bond checks & $z$-bond checks \\
\hline
$X_{1,i}X_{1,j},$ & $X_{1,i}^\dagger Z_{2,i}^\dagger X_{1,j}^\dagger Z_{2,j}^\dagger,$&$Z_{1,i}Z_{1,j},$\\
$X_{2,i}X_{2,j}$ & $ X_{2,i}^\dagger Z_{1,i}^\dagger X_{2,j}^\dagger Z_{1,j}^\dagger$&$Z_{2,i}Z_{2,j}$\\
 \hline
 \end{tabular}
\label{table: pauli schedule 2}
\end{table}

Anyons of $\TO$ are generated by $\F_i(e),\F_i(m)$, which have order $n$. Nontrivial statistics are: $\theta_{\F_1(e),\F_2(m)}=\theta_{e,m \sigma(\bar{m})=m^2}=2\pi \frac{2}{N}=\frac{2\pi}{n}$, and $\theta_{\F_1(m),\F_2(e)}=\theta_{m,e^2}=\frac{2\pi}{n}$ which corresponds to statistics of two copies of $\Z_n$ toric code if we  label the anyons as $e_1=\F_1(e),m_1=\F_2(m),e_2=\F_2(e),m_2=\F_1(m)$. The measurement induced automorphism, $\F_1(a)\leftrightarrow \F_2(a)$, is then: $\varphi: e_1\leftrightarrow e_2,m_1\leftrightarrow m_2$.
\subsection{$\TO_0 =\mathcal{D}_{\mathbb{Z}_N},\sigma: a\to \bar{a}, N=2n-1$}
The invariant subgroup is trivial. The onsite Hilbert space dimension is $D=|\TO_0|=N^2$.

$\mH=\mathbb{C}[\mathbb{Z}_N\times \mathbb{Z}_N]=\mathbb{C}_N\otimes \mathbb{C}_N$. A basis of $\mH$ is $|e^k,m^l\rangle$, the anyon braiding operators are 
\begin{align}
    T^e_z|e^k,m^l\rangle=e^{i\theta_{e,e^{2k}m^{2l}}}|e^k,m^l\rangle=e^{\frac{2\pi i 2l}{N}}|e^k,m^l\rangle
\end{align}
from which we read off $ T^e_z=Z_2^2$, similarly we get $T^e_x=X_1$, $T^m_z=Z_1^2$, $T^m_x=X_2$. $T^e_y=T_x^{e\dagger}T_z^{e\dagger}=X_1^\dagger Z_2^{\dagger 2}, T^m_y=T_x^{m\dagger}T_z^{m\dagger}=X_2^\dagger Z_1^{\dagger 2}$. \\

\begin{table}[H]
\centering
\setlength{\tabcolsep}{12pt} 
\renewcommand{\arraystretch}{2}
\begin{tabular}{|p{1.6cm}|p{2.2cm}|p{1.6cm}|}
\hline
$x$-bond checks & $y$-bond checks & $z$-bond checks \\
\hline
$X_{1,i}X_{1,j},$ & $X^\dagger_{2,i}Z^{\dagger 2}_{1,i}X^\dagger_{2,j}Z^{\dagger 2}_{1,j},$&$Z_{1,i}Z_{1,j},$\\
$X_{2,i}X_{2,j}$ & $X^\dagger_{1,i}Z^{\dagger 2}_{2,i}X^\dagger_{1,j}Z^{\dagger 2}_{2,j}$&$Z_{2,i}Z_{2,j}$\\
 \hline
 \end{tabular}
\label{table: pauli schedule 3}
\end{table}

In this case there are no invariant anyons.  Statistics of $\TO$ are: $\theta_{\F_1(e),\F_2(m)}=\theta_{e,m \sigma(\bar{m})=m^2}=2\pi \frac{2}{N}$ and $\theta_{\F_1(m),\F_2(e)}=\theta_{m,e^2}=2\pi\frac{2}{N}$, which give $\theta_{\F_1(e),\F_2(m)^M}=2\pi \frac{2M}{N}\equiv 2\pi \frac{1}{N} \mod 2\pi$ and similarly $\theta_{\F_1(m),\F_2(e)^M}=2\pi\frac{1}{N}$. Therefore we can label $e_1=\F_1(e),m_1=\F_2(m)^M,m_2=\F_1(m),e_2=\F_2(e)^M$ which form 2 copies of $\mathcal{D}_{\mathbb{Z}_N}$. The measurement induced automorphism is $\varphi:e_1=\F_1(e)\to \F_2(e)=((\F_2(e)^M)^2=e_2^2$ and $m_1=\F_2(m)^M\to \F_1(m)^M=m_2^M$.
\\

\subsection{$\TO_0 =\mathcal{D}_{\mathbb{Z}_N},\sigma: e\to m^p,m\to e^q,~pq~\mod N=1$}
$em^p$ is invariant: $em^p\to m^p (e^q)^p=m^pe$. The invariant subgroup is generated by $em^p$ and has order $N$.  Therefore the on-site Hilbert space dimension is $D=N^2/N=N$.

$\mH$ has a basis $|e\rangle, |e^2\rangle,\cdots, |e^N=1\rangle$. The anyon braiding operators are
\begin{align}
    T^e_z|e^k\rangle=e^{i\theta_{e,em^{-kp}}}|e^k\rangle=e^{\frac{2\pi i (-kp)}{N}}|e^k\rangle
\end{align}
from which we can read off $T^e_z=Z^{-p}$, similarly $T^e_x=X$, $T^e_y=T_x^{e\dagger}T^{e\dagger}_z=X^{\dagger}Z^p$.

\begin{table}[h]
\centering
\setlength{\tabcolsep}{12pt} 
\renewcommand{\arraystretch}{2}
\begin{tabular}{|M{2.0cm}|M{2.0cm}|M{2.0cm}|}
\hline
$x$-bond checks & $y$-bond checks & $z$-bond checks \\
\hline
$X_iX_j$ & $X_i^\dagger Z^p_iX_j^\dagger Z_j^p$&$Z_iZ_j$\\
 \hline
 \end{tabular}
\label{table: pauli schedule 4}
\end{table}

$\TO$ is generated by  anyons $\F_1(e),\F_2(e)$ with statistics $\theta_{\F_1(e),\F_2(e)}=\theta_{e,e m^{-p}}=2\pi \frac{-p}{N}$, which gives $\theta_{\F_1(e),\F_2(e)^{-q}}=2\pi \frac{1}{N}$, so we identify $\TO$ as $\mathcal{D}_{\mathbb{Z}_N}$ with $\F_1(e)$ being $e$ and $\F_2(e)^{-q}$ being $m$. The measurement induced automorphism is $\varphi: \F_1(e)\rightarrow \F_2(e)=(\F_2(e)^{-q})^{-p}$, i.e. $e\to \overline{m}^p$, similarly $m\to \overline{e}^q$. 

\end{document}